\documentclass[prb,twocolumn]{revtex4}
\usepackage{graphicx}

\newcommand{\sB}{{\scriptscriptstyle B}}
\newcommand{\sF}{{\scriptscriptstyle F}}
\newcommand{\sH}{{\scriptscriptstyle H}}
\newcommand{\sL}{{\scriptscriptstyle L}}
\newcommand{\sWL}{{\scriptscriptstyle WL}}

\begin{document}

\title{Two band transport and the question of a metal-insulator transition in
GaAs/GaAlAs two dimensional holes}

\date[]{September 25, 2001}

\author{Yuval Yaish}
\altaffiliation[Present address: ]{Laboratory of Atomic and Solid State
Physics, Cornell University, Ithaca, NY 14853.}
\author{Oleg Prus}
\author{Evgeny Buchstab}
\author{Gidi \surname{Ben Yoseph}}
\author{Uri Sivan}
\affiliation{Department of Physics and Solid State Institute, Technion-IIT,
Haifa 32000, Israel}
\author{Iddo Ussishkin}
\altaffiliation[Present address: ]{Department of Physics, Princeton
University, Princeton, NJ 08544.}
\author{Ady Stern}
\affiliation{Department of Condensed Matter Physics, The Weizmann Institute
of Science, Rehovot 76100, Israel}

\begin{abstract}
The magnetotransport of two dimensional holes in a GaAs/AlGaAs
heterostructure is studied experimentally and theoretically. Spin-orbit
splitting of the heavy hole band is manifested at high carrier densities in
two Shubnikov-de Haas frequencies, classical positive magnetoresistance,
and weak antilocalization. The latter effect combined with inelastic
scattering between the spin-orbit split bands lead to metallic
characteristics, namely resistance increase with temperature. At lower
densities, when splitting is smaller than the inverse elastic scattering
time, the two bands effectively merge to yield the expected insulating
characteristics and negative magnetoresistance due to weak localization and
interaction corrections to the conductivity. The ``metal to insulator''
transition at intermediate densities is found to be a smooth crossover
between the two regimes rather than a quantum phase transition. Two band
calculations of conventional interference and interaction effects account
well for the data in the whole parameter range.

\end{abstract}

\maketitle

\section{Introduction}
\label{Introduction}

The study of magnetotransport in two dimensional (2D) electronic systems
regained considerable interest since the observation of Kravchenko \emph{et
al.} \cite{Kravchenko94} that the resistance of high mobility, high density
silicon MOSFETs decreases and saturates to a residual value as the
temperature is reduced. This metallic characteristic is in sharp contrast
with the prevailing dogma that 2D systems are insulating \cite{gang4}
although interaction may possibly lead to delocalization. \cite{metal} The
conflict between Ref.~\onlinecite{Kravchenko94} and the expected insulating
characteristics motivated extensive experimental and theoretical efforts.
Soon, similar characteristics have been observed in other silicon samples,
\cite{silicon} SiGe quantum wells, \cite{SiGe} AlAs based 2D electron gas
(2DEG), \cite{AlAs} 2DEG in GaAs/AlGaAs heterostructures, \cite{2DEGGaAs}
2DEG in GaAs with self assembled InAs quantum dots, \cite{InAs} and various
realizations of 2D hole gas (2DHG) in GaAs/AlGaAs heterostructures.
\cite{Hanein1,Pepper1,Papadakis1,Sivan}

The samples that show metallic characteristics share some other features:

\noindent (a) Insulating behavior at low carrier densities, namely,
resistance increase with decreasing temperature.

\noindent (b) Weak dependance of the resistance upon temperature at some
intermediate density (coined ``metal-insulator transition'' (MIT)).

\noindent (c) Negative magnetoresistance at the density corresponding to
the MIT.

\noindent (d) Large, positive magnetoresistance for magnetic fields
parallel to the layer.

\noindent (e) A crossover from metallic to insulating characteristics for
large enough parallel magnetic fields.

The zero field crossover from metallic to insulating behavior was
identified by some authors as a second order phase transition.
\cite{superconductivity,Varma,Wignerglass} Scaling theory has been
constructed \cite{scaling} and even reentrant transition to an insulating
phase, at still higher densities, has been argued to occur for holes in
GaAs. \cite{Hamilton}

Notwithstanding the remarkable similarities between the magnetoresistance
and temperature/density dependences of the different material systems and
samples listed above, some gross differences should be appreciated:

\noindent (i) While the weak field magnetoresistance in silicon samples is
negative in all regimes, it is positive for holes in GaAs in most of the
metallic regime (high carrier concentration).

\noindent (ii) While only single band is observed in silicon samples, the
2DHG in GaAs samples, in most of the metallic regime, is characterized by
two distinct bands. These bands are manifested in two Shubnikov-de Haas
(SdH) frequencies, as well as in classical, positive magnetoresistance.

\noindent (iii) The relative resistance change with temperature in
different materials vary between about 1\% for electrons in GaAs to an
order of magnitude in silicon MOSFETs.

\noindent (iv) The ``critical resistance'' at the MIT varies substantially
from sample to sample and between different material systems. In contrast
to some early claims it can deviate substantially from one resistance
quantum \cite{Savchenko}.

\noindent (v) The large effect of back gating in GaAs 2DHG
\cite{Papadakis1,Papadakis2} is absent for electrons in silicon \cite{Prus}
or is at least very different \cite{Popovich}.

The extent to which the magnetotransport in different 2D systems is
universal is hence unclear at this point. The multivalley band structure of
silicon, the short range potential fluctuations in silicon as opposed to the
long range ones in GaAs, the strong spin-orbit coupling in III-V materials,
not to mention obvious differences in effective masses and Zeeman factors,
might all turn out to be important.

A wide spectrum of mechanisms has been proposed to account for these
features, including a new type of superconductivity,
\cite{superconductivity} a novel metallic phase induced by disorder
enhanced interactions \cite{Varma} (this direction is based on the earlier
works in Ref.~\onlinecite{metal}), Wigner glass and non-Fermi liquid,
\cite{Wignerglass} a new liquid phase, \cite{Newliquidphase} impurity
scattering, \cite{maslovaltshuler} temperature dependent screening,
\cite{Das Sarma,Hamilton2} spin effects, \cite{Popovich,Pudalov} interband
scattering, \cite{Sivan} band structure effects,
\cite{Papadakis1,Papadakis2} and classical percolation. \cite{Meir} None of
these explanations account for all features in all materials. It would thus
be fair to state that the metallic behavior, as well as the other features
listed above, remain unexplained as universal phenomena.

The present manuscript presents an extensive experimental and theoretical
study of magnetotransport in two dimensional hole gas in GaAs/GaAlAs
heterostructures. The experimental data pertain to measurements of high
mobility samples as well as low mobility ones. The data display all the
characteristic features listed above. At high densities all samples display
metallic characteristics
(Figs.~\ref{Fig3(high-mu,rxx,Q,S-vs.-T)},~\ref{Fig6(low-mu-rxx-vs.-T)}c-e).
At low densities the samples are insulating (Fig.~6a) in accordance with
point (a) above (we have chosen to display the insulating data for the low
mobility sample to avoid possible inhomogeneities that might occur at the
very low densities where the high mobility samples turn insulating). At an
intermediate density (Fig.~\ref{Fig6(low-mu-rxx-vs.-T)}b), the samples
display weak dependence of the resistance upon temperature (``MIT'', point
(b) above). All samples are characterized by a large, positive
magnetoresistance for a parallel magnetic field (points (d), (e) above and
Fig.~\ref{Fig7(low-mu-parallel-magnetic-field)}) and suppression of the
metallic characteristics by such a field.

In a recent letter \cite{Sivan} we reported magnetotransport measurements
done on a high mobility 2DHG sample. Concentrating on the high density
regime we were able to show that the metallic characteristics result from
inelastic scattering between the two heavy hole bands split by the
spin-orbit interaction. The positive magnetoresistance observed in that
sample was fully accounted for by the well known classical two band
formulae \cite{AshcroftMermin,Gantmakher,Zaremba} and the resistance
increase with temperature was shown to result from the enhancement of
Coulomb scattering between the two spin-orbit split bands. That work thus
explained the metallic characteristics with well known semiconductor
physics without invoking a novel metallic phase or any other ``new''
physics. The spin-orbit split bands were studied extensively both
experimentally, \cite{Stormer,Eisenstein,Shayegan,Papadakis1,Papadakis2}
and theoretically. \cite{BroidoSham,Ekenberg,Goldoni,Winkler} In fact,
Murzin \emph{et al.} \cite{Murzin} have previously shown that interband
Coulomb scattering is manifested in a resistance increase with temperature.

The present manuscript extends our previous work in several ways. First we
refine the high density data analysis and show that the negative
magnetoresistance at larger magnetic fields results from quantum
corrections due to Coulomb interaction. \cite{EEIreview,Paalanen,Choi} Next
we report new measurements covering the insulating, MIT, and metallic (high
density) regimes. The data conform with previous magnetotransport
measurements on 2DHG.
\cite{Hanein1,Pepper1,Papadakis1,Papadakis2,Murzin,Eisenstein,Hanein2,Pepper2}
Careful analysis then shows that well-known physics accounts for all data.
The emerging picture is briefly as follows.

It is well known, both theoretically
\cite{BroidoSham,Ekenberg,Goldoni,Winkler} and experimentally
\cite{Papadakis1,Sivan,Papadakis2,Stormer,Eisenstein,Shayegan}, that the
bulk heavy hole band in assymmetric GaAs/AlGaAs heterostructures is split
by spin-orbit interaction. The resulting sub-bands are manifested by two
SdH frequencies, as well as by a classical, Lorentzian shaped, positive
magnetoresistance (PMR). The band splitting energy, $\epsilon_g$, vanishes
at $k = 0$ and increases with $k$. \cite{BroidoSham} At low densities, when
$\epsilon_g$ at the Fermi energy is small and the elastic scattering time,
$\tau$, is short, $\epsilon_g \ll \hbar / \tau$, the two bands are strongly
mixed and spin-orbit splitting is unimportant. The resulting
magnetoresistance is negative and dominated by weak localization (WL) and
hole-hole Coulomb interaction. As the temperature is increased, these
quantum corrections are suppressed and the sample becomes more conductive.
At high densities, the sub-band splitting is large and $\tau$ is long, thus
$\epsilon_g \gg \hbar / \tau$. The two sub-bands are hence distinguishable.
The weak field magnetoresistance is positive and dominated by weak
anti-localization (WAL) induced by inter sub-band scattering (effectively,
spin-orbit scattering), hole-hole interaction (Altshuler-Aronov correction
\cite{EEIreview}), and classical two band magnetoresistance. A slightly
higher magnetic field suppresses the quantum corrections in the Cooperon
channel while the classical magnetoresistance persists. At still higher
fields, the classical magnetoresistance is also suppressed and the
magnetoresistance turns negative due to hole-hole interaction in the
diffuson channel.

The metallic characteristic, $d \rho / d T > 0$, in the high density regime
is traced to the enhancement of inelastic inter sub-band scattering as well
as the suppression of quantum corrections (especially the WAL) with
increased temperature. The weight of quantum corrections relative to the
classical effect depends on sample resistivity. The quantum corrections are
negligible in high mobility samples and very important in the low
mobility ones  close to the MIT. The weak temperature dependence of the
resistivity at the MIT merely reflects effective cancellation of the
resistance increase with temperature due to inelastic interband scattering
and suppression of WAL, against resistance decrease with temperature due to
the suppression of hole-hole interaction.

Our analysis thus explains how the sample crosses over from an insulator at
low densities to a ``metal'' at higher densities. Moreover, we account for
the change from negative to positive magnetoresistance as the density is
increased. All these features emerge from the well known band structure of
holes combined with conventional quantum corrections to the
magnetoresistance. No new metallic phase is invoked.

The manuscript comprises the following parts. Section \ref{High mobility}
provides a refined analysis of our previously published data \cite{Sivan}
on high mobility samples. The refined analysis includes hole-hole
interaction in the diffuson channel. The extracted inter and intraband
scattering rates remain unchanged. The analysis provides new information
about the Coulomb interaction parameter at different densities. The theory
of quantum corrections to the magnetotransport is briefly reviewed in
Section \ref{High mobility B}. Previous calculations are extended to
account for two sub-bands with different mobilities. Details of these
calculations are provided in the Appendix. Extensive data measured on a
lower mobility sample are provided in section \ref{Low mobility}. The data
are analyzed taking into account WL, WAL, hole-hole interaction in the
cooperon and diffuson channels, and interband scattering. A comprehensive
picture of magnetotransport in a 2DHG is compiled in section
\ref{Discussion}. Section \ref{Calculation} details some theoretical
aspects of magnetotransport in a two band system with spin-orbit splitting.
Plasmon mediated Coulomb scattering is analyzed in this section, and shown
to yield an Arrhenius temperature dependence. Finally, this section also
discusses the quantum corrections to the magneto-conductivity in this
system.

\section{High mobility 2DHG sample and theoretical background}
\label{High mobility}

\subsection{Experiment}
\label{High mobility A}

The high mobility sample had a mobility of $\simeq 300,000 \text{~cm}^2
\text{/(V\,s)}$ at $p = 4 \times 10^{11} \text{~cm}^{-2}$ and $T = 200
\text{~mK}$. The low mobility sample had a $\simeq 20,000 \text{~cm}^{2}
\text{/(V\,s)}$ mobility at a similar carrier density and $T = 400
\text{~mK}$. The 2DHG was confined in both samples to a
GaAs/Al$_{0.8}$Ga$_{0.2}$As interface in the $\langle 100 \rangle$ plane.
The samples had a 2DEG front gate, 40 and 20~nm (high and low mobility
samples, respectively) above the 2DHG and a silicon doped back gate, 300~nm
below it \cite{Sivan}.

\begin{figure}
\begin{center}
\includegraphics[width=3.4in]{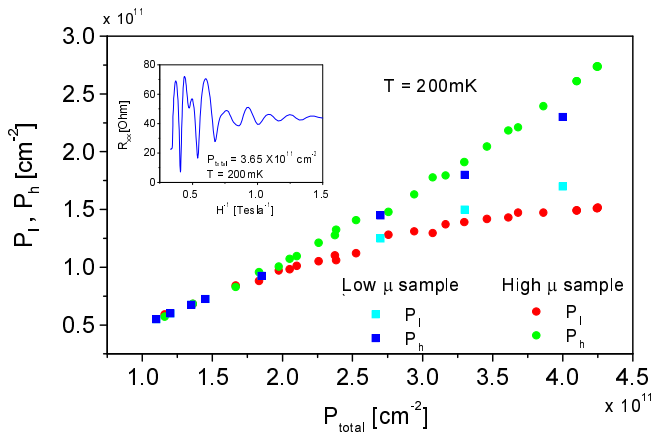}
\end{center}
\caption{Hole density in the light $\left( p_{l}\right) $ and heavy $\left(
p_{h}\right) $ spin-orbit split bands as a function of total density. Inset
- One of the Shubnikov de Haas traces (high mobility sample) used to
determine the hole densities.} \label{Fig1(populations)}
\end{figure}
The inset to Fig.~\ref{Fig1(populations)} depicts a characteristic SdH
curve. Two distinct frequencies corresponding to two bands are clearly
observed and used to determine the carrier densities in the lighter and
heavier bands ($p_l$ and $p_h$, respectively, not to be confused with the
bulk light and heavy bands). Fig.~\ref{Fig1(populations)} depicts $p_l$ and
$p_h$ as a function of the total density, $p_{total}$, for the high
(circles) and low (squares) mobility samples. The band's population in the
two samples are similar. Below a total density of about $2 \times 10^{11}
\text{~cm}^{-2}$ it is hard to resolve two bands in the SdH data. For
higher densities, the bands split and for $p_{total} \geq 4 \times 10^{11}
\text{~cm}^{-2}$  practically all additional carriers populate the heavier,
less mobile, band.

\begin{figure}
\setlength{\unitlength}{3.4in}
\begin{picture}(1,0.7)
\put(0.08,0){\includegraphics[width=0.92\unitlength]{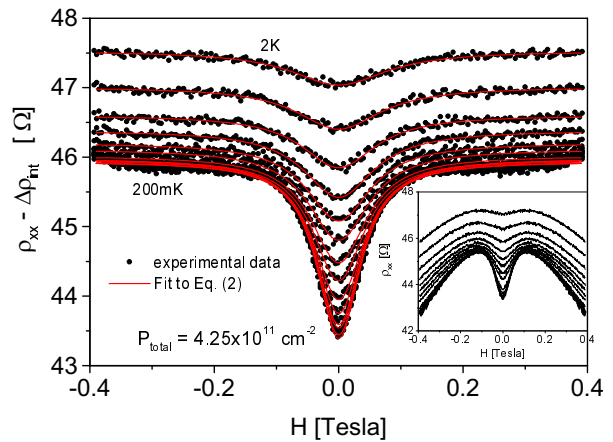}}
\end{picture}
\caption{The two-band classical contribution to the magnetoresistance  for
different temperatures (high mobility sample). Note the perfect Lorentzian
shape MR at low fields. Inset- Full magnetoresistance curves for the same
temperatures.} \label{Fig2(high-mu-MR)}
\end{figure}
The weak field MR of the high mobility sample at $p_{total} = 4.25 \times
10^{-11} \text{~cm}^{-2}$, $p_l = 1.52 \times 10^{-11} \text{~cm}^{-2}$,
$p_h = 2.73 \times 10^{-11} \text{~cm}^{-2}$, and different temperatures is
presented in the inset to Fig.~\ref{Fig2(high-mu-MR)}. The Lorentzian
shaped PMR expected from two band transport is obtained by subtracting the
predicted quantum correction (see below) from the full MR data. The result
is depicted in Fig.~\ref{Fig2(high-mu-MR)}.

The analysis of the high mobility data presented in \cite{Sivan} was based
on standard, two-band transport formulae, \cite{AshcroftMermin} generalized
to include interband scattering. \cite{Gantmakher,Zaremba} The Lorentzian
PMR expected for two band transport was extracted by subtracting the weak
parabolic negative magnetoresistance (NMR), attributed to Coulomb
interactions \cite{Paalanen}, from the full MR curve. Here we refine the
analysis and incorporate quantum corrections \emph{ab initio}. These
corrections include hole-hole interaction, WL, and WAL. They all become
increasingly important as the sample conductivity is reduced. Their
inclusion \emph{ab initio} facilitates a unified analysis of high and low
mobility samples. Moreover, the quantum corrections provide valuable
information on the fundamental scattering processes in 2DHG.

\subsection{Theoretical background}

\label{High mobility B}

Our analysis generalizes Choi \emph{et al.} \cite{Choi} approach to the
case of two band transport. The starting point is a $4 \times 4$
conductivity matrix for the two bands in the presence of a perpendicular
magnetic field. All matrix elements, except the Hall ones, are modified by
quantum effects. Setting the Hall currents to zero and inverting the matrix
one obtains the longitudinal resistance with WL, WAL, and interaction
corrections:
\begin{equation} \label{MR(eq.)}
\rho_{xx} (H) = \rho_\sL (H)+ f(S_{l}, S_{h}, Q, H) \cdot (\delta
\sigma_\sWL + \delta \sigma_{int}) \, ,
\end{equation}
where
\begin{eqnarray} \label{Lorentzian (eq.)}
\rho_\sL (H) & = & \rho_\sL (H \rightarrow \infty) + \frac{L}{1 +
(H/W)^{2}} \, , \nonumber \\ \rho_\sL (H \rightarrow \infty) & = &
\frac{R_h^2 S_l + R_l^2 S_h - 2 Q R_l R_h}{(R_l + R_h)^2} \, ,  \\
\nonumber W & = & \frac{S_l + S_h + 2 Q}{R_l + R_h} \, , \\
\nonumber L & = & - \frac{\left[ R_l (S_h + Q) - R_h (S_l + Q)
\right]^2}{(S_l + S_h + 2 Q) (R_l + R_h)^2} \, ,
\end{eqnarray}
and
\begin{eqnarray*}
\delta \sigma_\sWL & = & \delta \sigma_\sWL^{ll} + \delta \sigma_\sWL^{lh}
+ \delta \sigma_\sWL^{hl} + \delta \sigma_\sWL^{hh} \, , \\
\delta \sigma _{int} & = &\delta \sigma_{int}^{ll} + \delta
\sigma_{int}^{lh} + \delta \sigma_{int}^{hl} + \delta \sigma_{int}^{hh} \,
.
\end{eqnarray*}
Here, $l,h$ correspond to the lighter and heavier heavy hole sub-bands,
respectively. $R_i = 1 / ep_i$, with $i=l,h$, is the Hall coefficient of
the $i$-th band. The diagonal resistances $S_l$, $S_h$, and off-diagonal
resistance $Q$ can be expressed in terms of elastic and inelastic
contributions, $S_l (T) = S_l (0) + \alpha^{-1} [Q (T) - Q (0)]$; $S_h (T)
= S_h (0) + \alpha [Q (T) - Q (0)]$, where $\alpha$ is a function of
velocities and densities. The resistances $S_l (0)$, $S_h (0)$ pertain to
inter and intraband impurity scattering, $Q (0)$ and $Q (T) - Q (0)$
pertain to elastic and inelastic scattering, respectively. The latter
processes may include carrier transfer between bands as well as drag-like
processes where a particle from one band scatters off a particle in the
other band and both carriers maintain their bands. The resistance
$\rho_\sL$ reflects classical, two band PMR and depends in a Lorentzian way
on the magnetic field. The conductances, $\delta \sigma_\sWL$ and $\delta
\sigma_{int}$, stand for WL (or WAL) and interaction corrections to the
conductivity. The function $f (S_{l}, S_{h}, Q, H)$ given in the appendix,
is roughly quadratic in $H$. It is negative for $H < W$ and positive for $H
> W$ (see Fig.~\ref{fig20(f(S_l,S_h,Q,H) vs. H)}). The functions $\delta
\sigma_\sWL$, $\delta \sigma_{int}$, and $f (S_{l}, S_{h}, Q, H)$ depend on
both temperature and the magnetic field. Equation~(\ref{MR(eq.)}) shows
that the overall quantum contribution to the resistance comprises the
corrections to the conductivity of both bands. A qualitative understanding
of $f$ can be gained from the single band case where it reduces to $-
\rho_0^2 \left[ 1 - (\omega_c \tau)^2 \right]$, with $\omega_c = e H / m c$
being the cyclotron frequency [cf.\ Eq.~(\ref{one-band-MR(eq.)})]. The
quadratic increase of $f$ with $H$ yields a large interaction correction to
the magneto-resistance despite the smallness of $\delta \sigma_{int}$.

We turn now to review the theory for the quantum corrections, WL, WAL, and
interaction, taken into account in the data analysis. Further details are
provided in Sec.~\ref{Calculation}. We start with WL and WAL. Various
authors \cite{BroidoSham,Ekenberg,Goldoni,Winkler} calculated the hole band
structure of asymmetric GaAs/AlGaAs quantum wells and heterostructures and
found that each band is characterized by a wave-vector dependent spinor
state. When a hole is scattered from $\mathbf{k}$ to $\mathbf{k'}$ the
spin-orbit coupling leads to Dyakonov-Perel spin precession around
$\mathbf{k'}\,$ \cite{Dyakonov,Dresselhaus,Knap,Lyanda-Geller,Averkiev1}.
As a result, the spin relaxes at a rate $\hbar / \tau_{so} \approx
\epsilon_g^2 \tau / \hbar$, where $\tau$ is the momentum relaxation time,
and $\epsilon_g$ is the energy difference between the two bands for an
average $k_\sF$. Since WAL is important for $g \approx 1$, where the
momentum relaxation rate is similar to the single particle scattering time,
we neglect the difference between the two. Scattering between two
spin-orbit split bands is hence equivalent to spin-orbit scattering by,
e.g., a large atom. For $\tau_{so} < \tau_\varphi$ ($\tau_\varphi$ is the
dephasing time), WAL takes place with either positive or negative MR for
$l_\sH = \sqrt{\hbar c / e H}$ larger or smaller than $l_{so} = \sqrt{D
\tau _{so}}$ ($D$ is the diffusion coefficient). Averkiev \emph{et al.}
\cite{Averkiev1} calculated the WL correction for weak and strong
spin-orbit interaction in symmetric p-type quantum wells. They found that
the key parameter for the anomalous MR is $k_\sF a / \pi $, where $a$ is
the well width. This parameter is a measure of bulk heavy/light hole wave
function mixing. For $k_\sF a / \pi \sim 1$ they predict PMR at weak fields
and NMR at higher fields. Pedersen \emph{et al.} \cite{Pedersen1} measured
such effect in a symmetric GaAs p-type quantum well. In section
\ref{6:QuantumSec} we generalize Averkiev \emph{et al.} procedure to the
case of asymmetric GaAs p-type quantum wells with their spin-orbit split
bands, taking into account interband and intraband particle-particle
propagator contributions to the Cooperon equation. The final result is
\begin{eqnarray}\label{WALformula}
\lefteqn{\delta \sigma_\sWL (H) \equiv \sigma_\sWL (H) - \sigma_0 (H) =
-\frac{e^2}{4 \pi^2 \hbar}} \\ \nonumber \\ & \times & \left[ 2 \Psi \left(
\frac{1}{2} + \frac{H_{tr}}{H} \right) - 2 \Psi \left( \frac{1}{2}
+ \frac{\frac{1}{2} H_{so} + H_\varphi}{H} \right) \right. \nonumber \\
\nonumber && \;\;\;\;\;\;\;\;\;\;\;\;\;\;\;\; \left. -\Psi \left(
\frac{1}{2} + \frac{H_{so} + H_\varphi}{H} \right) + \Psi \left(
\frac{1}{2} + \frac{H_\varphi}{H} \right) \right] \, ,
\end{eqnarray}
where $\Psi$ is the Digamma function, and the characteristic magnetic
fields $H_{tr}$, $H_{so}$ and $H_\varphi$ are given by
\[
H_{tr} = \frac{\hbar c}{4 e D \tau} \, , \qquad H_{so}=\frac{\hbar c}{4 e D
\tau_{so}} \, , \qquad H_\varphi = \frac{\hbar c}{4 e D \tau_\varphi} \, .
\]
Using Dyakonov-Perel equation, $\hbar / \tau_{so} \approx \epsilon_g^2 \tau
/ \hbar$, it is then possible to estimate the spin-orbit band splitting,
$\epsilon_g$, from transport measurements. The existence of two bands is
manifested in three phenomena; a Lorentzian PMR, two SdH frequencies, and
finally, WAL correction to the conductivity. While the SdH frequencies
merge into a single frequency when the gap becomes smaller than the
scattering rate ($\epsilon_g < \hbar /\tau$), the WAL remains visible down
to considerably smaller gaps provided the temperature is low enough. The
WAL is hence a powerful tool in the analysis of the 2DHG band structure and
scattering mechanisms. The WAL is visible as long as $\tau_{so} \leq
\tau_\varphi$. Substituting $\hbar / \tau_\varphi \simeq T / g$ for the
dephasing rate we find WAL should be visible as long as $T < g
\frac{\hbar}{\tau} \left( \frac{\epsilon_g \tau}{\hbar} \right)^{2}$,
although $\epsilon_g \tau / \hbar$ may be smaller than unity. Moreover,
since the relevant magnetic field for WAL is $g$ times smaller than that
relevant for the classical effect or the SdH oscillations, the WAL probes
the two bands at very small magnetic fields. Winkler \emph{et al.}
\cite{Winkler} argue that SdH data at finite fields underestimates the zero
field band splitting. Our analysis proposes WAL as a better tool for
studying degeneracy lifting of the heavy holes, particularly at weak
magnetic fields. The WL correction is significant as long as $l_\sH \geq l$
or $\sqrt{2} \omega _{c} \tau g \leq 1$, where $\omega_c = e H / m c$ is
the cyclotron frequency and $l = \sqrt{D \tau}$.

We turn now to the interaction corrections \cite{EEIreview} to the
conductivity. Our analysis follows Choi \emph{et al.} \cite{Choi} Note the
interaction MR is mainly due to the magnetic field dependence of the
pre-factor of these corrections, $f (S_l, S_h, Q, H)$ in
Eq.~(\ref{MR(eq.)}), rather than the explicit dependence upon $H$ expressed
in Eqs.~(\ref{Diffusion-correction(eq.)}) and~(\ref{F(eq.)}).

The diffuson channel correction is
\cite{Fukuyama2,Finkel'stein,Fukuyama1,Lee}
\begin{eqnarray}\label{Diffusion-correction(eq.)}
\lefteqn{\delta \sigma _{int}^{D}(T,H)=} \\ \nonumber & & \!\!\!\!\!\!\!\!
\left\{
\begin{array}{ll}
\!\! -\frac{e^{2}}{2\pi ^{2}\hbar }g_{0} \ln \left( \frac{\hbar }{T \tau}
\right) -\frac{e^{2}\Xi }{4\pi ^{2}\hbar }g_{2}(h), & \text{for }
\frac{\hbar }{T \tau} \! > \! 1 \\ \\ \!\! \frac{e^{2}}{2\pi ^{2}\hbar
}g_{0} \! \left[ \Psi \! \left( \frac{1}{2} \! + \! \frac{\hbar }{T
\tau}\right) \! - \! \Psi \! \left( \frac{1}{2}\right) \right] \! - \!
\frac{ e^{2}\Xi }{4\pi ^{2}\hbar }g_{2}(h), & \text{for }\frac{ \hbar }{T
\tau} \! < \! 1
\end{array}
\right.
\end{eqnarray}
where
\begin{eqnarray}
g_{0} & = & 4-3\frac{2+F}{F} \ln \left( 1+\frac{F}{2}\right) ,  \nonumber \\
\Xi & = & 4\left[ \frac{2+F}{F}\right] \ln (1+\frac{F}{2})-4,
\label{F(eq.)} \nonumber \\ F & = &\int_{0}^{2\pi }\frac{d\theta }{2\pi
}\left[ 1+\frac{2K_{F}}{\kappa
}\sin \left( \frac{\theta }{2}\right) \right] ^{-1}, \\
g_{2}(h) & =& \int_{0}^{\infty }d\Omega \frac{d^{2}}{d\Omega ^{2}}\left[
\Omega \nu (\Omega )\right] \ln \left| 1-\frac{h^{2}}{\Omega ^{2}}\right| ,
\nonumber \\
h & = & g^{\star }\mu _{B}H/T,  \nonumber
\end{eqnarray}
and
\[
g_{2}(h)=\left\{
\begin{array}{ll}
\ln (h/1.3) & \text{for }h\gg 1, \\ 0.084h^{2} & \text{for }h\ll 1.
\end{array}
\right.
\]
$\Psi$ is the Digamma function, $\kappa$ is the 2D inverse screening
length, $\nu (\Omega)$ is the density of states at an energy $\hbar
\Omega$, $g^\star$ is the 2DHG effective g-factor, and $\mu_\sB$ is the
Bohr magneton. In the diffuson channel the interaction correction depends
on the magnetic field through the Zeeman energy only. The diffusive
correction, $\delta \sigma_{int}^D$, is hence independent of $H$ as long as
$h \ll 1$. Unlike the cooperon correction that vanishes for $\omega_c \tau
> 1 / g$, the diffuson correction persists to $\omega_c \tau \simeq 1$.

The cooperon channel correction to the conductivity is given by
\cite{Altshuler}
\begin{eqnarray*}
\delta \sigma _{int}^{C}(T,H) &=&-\frac{e^{2}}{2\pi ^{2}\hbar }
g_{c}(T,H) \ln \left( \frac{T \tau}{\hbar }\right) \\
&&-\frac{e^{2}}{2\pi ^{2}\hbar }g_{c}(T,H)\phi _{2}\left( \frac{2eDH}{\pi T
c} \right) ,
\end{eqnarray*}
where
\begin{eqnarray*}
\phi _{2}(x) &=&\int_{0}^{\infty }\frac{t dt}{\sinh ^{2}(t)}\left[ 1 -
\frac{xt}{\sinh (xt)} \right] , \\
\phi _{2}(x) &=&\left\{
\begin{array}{ll}
\ln (x) & \text{for }x\gg 1, \\ 0.30x^{2} & \text{for }x\ll 1,
\end{array}
\right.
\end{eqnarray*}
and
\[
g_{c}^{-1}(T,H)=\left\{
\begin{array}{ll}
g_{0}^{-1}+\ln \left( \frac{1.13E_{F}}{T} \right) & \text{for }\frac{2eDH}{
\pi Tc}<1,
\\ \\ g_{0}^{-1}+\ln \left( \frac{cE_{F}}{DeH}\right) & \text{for }
\frac{2eDH}{\pi Tc}>1.
\end{array}
\right.
\]

\noindent The correction depends on $H$ and vanishes for $H>2\pi H_{tr}$,
or equivalently, $\sqrt{2}\omega _{c} \tau g/2\pi >1$. The Maki-Thompson
correction is ignored.

\subsection{Data analysis}
\label{High mobility C}

\begin{figure}
\begin{center}
\setlength{\unitlength}{3.4in}
\begin{picture}(1,0.68)
\put(0.05,0){\includegraphics[width=0.95\unitlength]{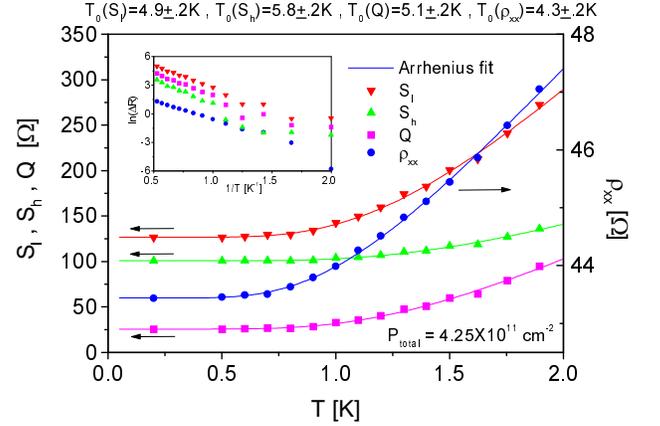}}
\end{picture}
\end{center}
\caption{The various scattering rates expressed as resistances (left axis)
and the zero field longitudinal resistance (right axis) vs. $T$. Solid
lines depict best fit to Arrhenius dependence with the characteristic
temperatures listed at the top of the figure. Inset - same data in semi log
plot.} \label{Fig3(high-mu,rxx,Q,S-vs.-T)}
\end{figure}
For the high mobility sample, the WL and WAL corrections are negligible
above 1 Gauss. The analysis may thus be confined to interband scattering
and interaction in the diffuson channel
[Eq.~(\ref{Diffusion-correction(eq.)}) \cite{Fukuyama2}]. The full MR is
depicted in the inset to Fig.~\ref{Fig2(high-mu-MR)}. The parabolic
background is attributed to Coulomb interaction and analyzed below.
Subtracting this background we obtain the data presented in
Fig.~\ref{Fig2(high-mu-MR)}. Fitting these data to Eqs.~(\ref{Lorentzian
(eq.)}), one obtains $\rho_{xx} (H \rightarrow \infty)$, $L$, and $W$ as a
function of temperature and density. Note the excellent agreement with the
predicted shape, Eqs.~(\ref{Lorentzian (eq.)}). Below $\simeq 0.6
\text{~K}$, the resistance is practically independent of $T$ while for
temperatures above $\simeq 2 \text{~K}$, the Lorentzian is hardly visible.
The suppression of the classical two band magnetoresistance results from
interband scattering. At low temperatures this scattering is mainly
elastic. As the temperature is increased, inelastic scattering commences,
the drift velocities of carriers in the two bands gradually approach each
other, and the magnetoresistance is consequently diminished. The extracted
functions, $S_l$, $S_h$, and $Q$, together with $\rho_{xx} (H = 0)$, are
shown vs.\ $T$ in Fig.~\ref{Fig3(high-mu,rxx,Q,S-vs.-T)} for the same total
density as in Fig.~\ref{Fig2(high-mu-MR)}. At low temperatures, all these
quantities saturate to some residual values $S_l(0)$, $S_h(0)$, $Q(0)$
which we attribute to inter and intraband elastic scattering. As the
temperature is increased, inelastic scattering commences and these
quantities grow.

\begin{figure}
\begin{center}
\includegraphics[width=3.4in]{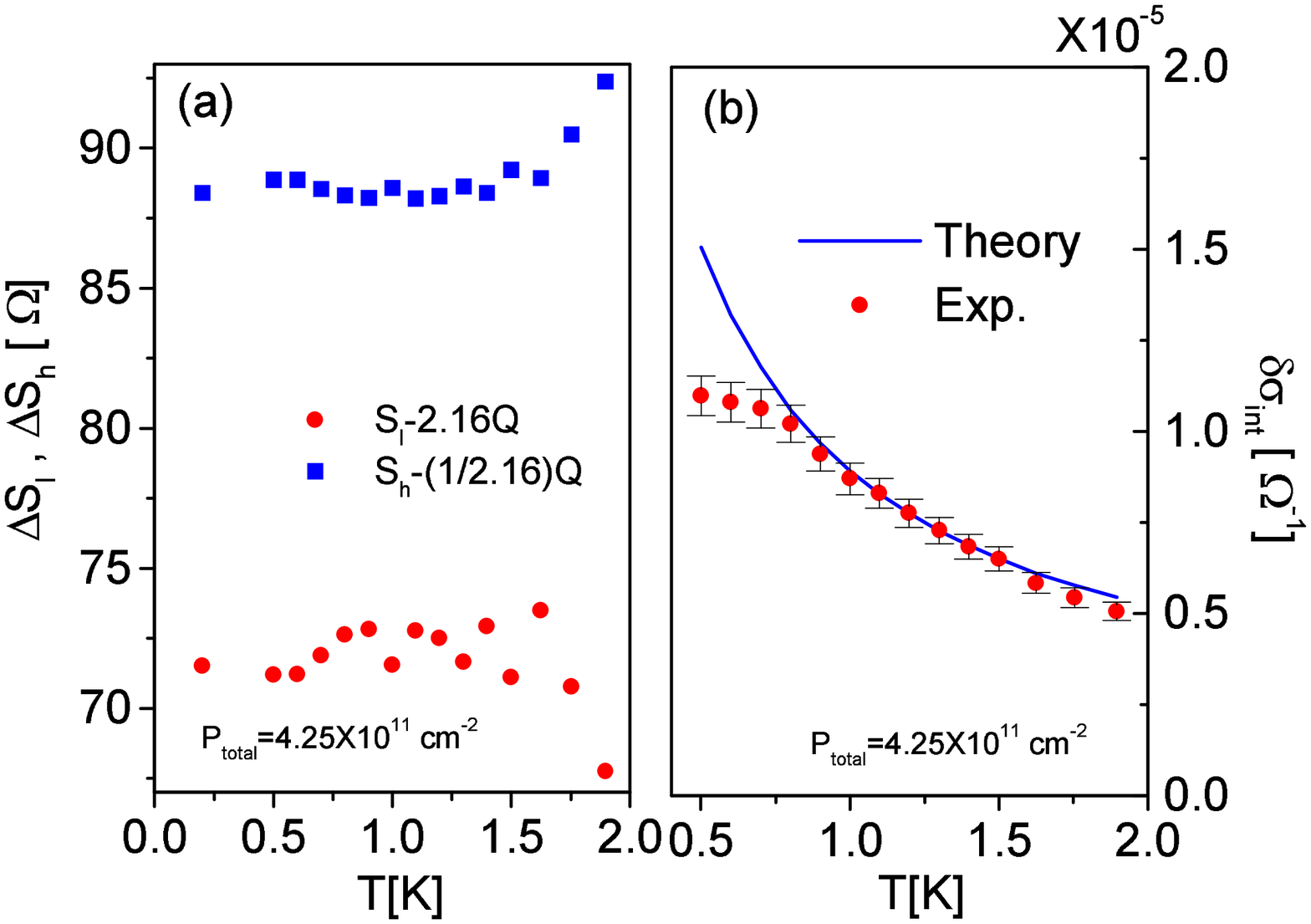}
\end{center}
\caption{(a) High mobility sample. $\Delta S_{l}=S_{l}-2.16Q$, $\Delta
S_{h}=S_{h}-2.16^{-1}Q$ vs. $T$ for $p=2\times 10^{11}cm^{-2}$. (b)
Hole-hole interaction correction to the conductance in the diffuson
channel, $\protect\delta \protect\sigma _{int}^{D}$.}
\label{Fig4(high-mu-a)S-2.16Q-vs.-T,b)Qcorrection-vs.-T)}
\end{figure}
The remarkable and central result emerging from the high mobility data is
the observation that the inelastic scattering rates follow the same
temperature dependence as $\rho_{xx}(H = 0)$, namely, $S_{i} (T) = S_{i}
(0) + \alpha_{S_i} \exp (-T_0 / T)$, $Q (T) = Q (0) + \alpha_Q \exp (-T_0 /
T)$, where $\alpha_{S_i}$, $\alpha_Q$ are constants. The characteristic
temperature, $T_0$, is similar to all these quantities, including
$\rho_{xx}$ (see inset to Fig.~\ref{Fig3(high-mu,rxx,Q,S-vs.-T)}). For
$S_l$ we find $T_0 = 4.9 \pm 0.2$~K, for $S_h$ we find $T_0 = 5.8 \pm
0.2$~K, for $Q$ we find $T_0 = 5.1 \pm 0.1$~K, and for $\rho_{xx}$ we find
$T_0 = 4.3 \pm 0.1$~K. We find experimentally [see
Fig.~\ref{Fig4(high-mu-a)S-2.16Q-vs.-T,b)Qcorrection-vs.-T)}(a)] that for
$T < 1.6$~K, $S_l (T) \simeq S_l (0) + 0.46^{-1} [Q(T) - Q(0)]$; $S_h (T)
\simeq S_h (0) + 0.46 [Q(T) - Q(0)]$, thus yielding for the density of
Figs.~\ref{Fig2(high-mu-MR)} and~\ref{Fig3(high-mu,rxx,Q,S-vs.-T)}, $\alpha
= 0.46$. Since the pre-factors multiplying $Q$ are reciprocal, the
resistances $S_l (0)$ and $S_h(0)$ are identified as the diagonal
resistances, pertaining to intraband scattering. The Arrhenius temperature
dependence is hence traced to inter-band scattering alone, rendering the
intra-band scattering practically temperature independent. We attribute
this unexpected scattering rate to the acoustic plasmon mediated Coulomb
interaction discussed below. The results are identical to those presented
in Ref.~\onlinecite{Sivan}.

We turn now to analyze the interaction correction to the conductivity. The
Coulomb interaction is characterized by $F$, the angular average of the
statically screened Coulomb interaction, given by Eq.~(\ref{F(eq.)}). With
$S_l, S_h, Q$ known from the previous fit, we fit the parabolic background
to Eq. (\ref{MR(eq.)}) to extract $\delta \sigma _{int}$. We find $F = 0.5
\pm .05$ for all $T$ except the lowest ones. This value should be compared
with the theoretical value, $F = 0.81$.
Fig.~\ref{Fig4(high-mu-a)S-2.16Q-vs.-T,b)Qcorrection-vs.-T)}b compares the
experimental $\delta \sigma _{int}$ with the theoretical one as a function
of $T$. For temperatures above $0.8 \text{~K}$, the agreement between
theory and experiment is excellent. At lower temperatures the experimental
interaction correction is smaller than that predicted by theory. The
interaction coefficient, $g_{0}(F = 0.47)$ [Eq.~(\ref{F(eq.)})], is $0.65
\pm 0.03$, compared with $g_{0}(F = 0.81) = 0.46$, predicted by theory. It
should be noted that the theoretical value is calculated for two identical
parabolic bands with $m=0.3m_{0}$. The actual band structure is more
complex and in fact, non-parabolic. The fact that $F$ is independent of
temperature above $0.8K$ supports the attribution of the interaction
contribution to the diffuson channel. We emphasize that the saturation of
the interaction correction below $0.8K$ is real and does not result from
carrier heating. The latter, deduced from the SdH data, proves efficient
cooling down to $0.2K$.

\begin{figure}
\begin{center}
\setlength{\unitlength}{3.4in}
\begin{picture}(1,0.7)
\put(0.06,0){\includegraphics[width=0.94\unitlength]{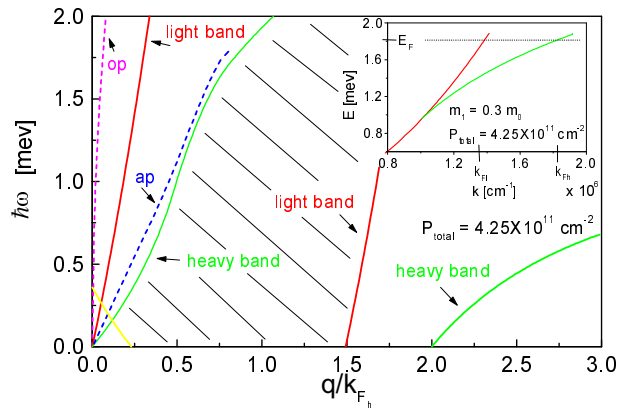}}
\end{picture}
\end{center}
\caption{Solid lines - heavy and light particle-hole excitation continua as
a function of momentum scaled to the heavy hole Fermi wave vector. Shaded
area corresponds to the range where drag-like interband scattering is
possible at very low $T$. Dashed lines -optical (op) and acoustic (ap)
plasmon dispersions. Inset - The measured bands dispersion relations. The
energy $E_{F}$ corresponds to the hole Fermi energy at $p_{total} = 4.25
\times 10^{11} \text{~cm}^{-2}$.} \label{Fig5(e-h-continuum)}
\end{figure}
In section \ref{Calculation} we discuss in detail possible reasons for the
Arrhenius temperature dependence of $S_{l},S_{h}$ and $Q$ which in turn
leads to a similar temperature dependence of $\rho _{xx}$. These scattering
rates (expressed as resistances) crucially depend on the bands' dispersions
relations, $E_{i}({\bf k})$, and their resulting excitation spectra. To
extract the bands dispersions depicted in the inset to
fig.~\ref{Fig5(e-h-continuum)}, we approximate \cite{BroidoSham,Ekenberg}
the light band by a parabolic relation with a mass $m_{l}=0.28 m_{0}$
($m_{0}$ is the bare electron mass) which was found from SdH temperature
dependence. The variation of $p_{l}$, $p_{h}$ with $p_{total}$, depicted in
fig.~\ref{Fig1(populations)}, is then used to calculate the ratio between
the two bands compressibilities. Neglecting band warping as well as
differences between density of states and compressibility, we use the ratio
of the two compressibilities to extract the dispersion of the heavy band.
This dispersion then allows, within the random phase approximation, the
calculation of the excitation spectrum of the system. The spectrum is
composed of two particle-hole continua, one for each band, and two plasmon
branches. Both are shown in fig.~\ref{Fig5(e-h-continuum)} for zero
temperature.

In summary, we find that the MR of high mobility 2DHG is governed by
classical, two band PMR at weak fields and NMR (parabolic background) due
to Coulomb interaction in the diffuson channel at stronger fields. The
resistance increase with temperature (``metallic phase'') follows the
inelastic interband scattering. Careful analysis of the interaction
contribution to the conductance reveals unexpected saturation of the
interaction below $0.8K$.

\section{Low Mobility 2DHG SAMPLE}
\label{Low mobility}

\begin{figure}
\begin{center}
\includegraphics[width=3.4in]{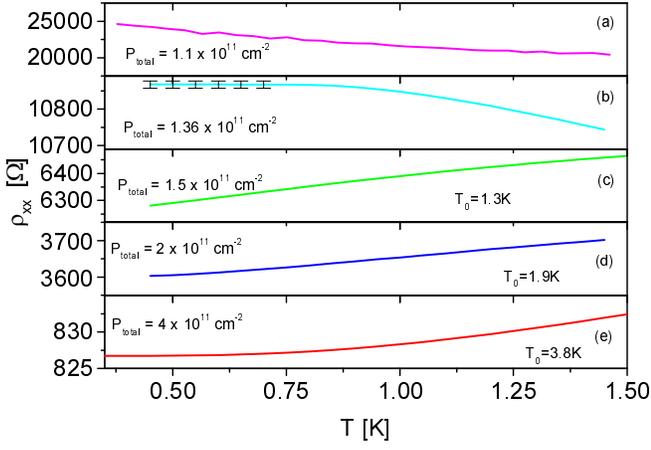}
\end{center}
\caption{Low mobility sample. Longitudinal resistance vs. temperature at
different carrier densities. In the metallic regime the data are fitted to
an Arrhenius function. The characteristic temperature, $T_{0}$, is
indicated in the figure.} \label{Fig6(low-mu-rxx-vs.-T)}
\end{figure}
\begin{figure}
\begin{center}
\setlength{\unitlength}{3.4in}
\begin{picture}(1,0.64)
\put(0.08,0){\includegraphics[width=0.92\unitlength]{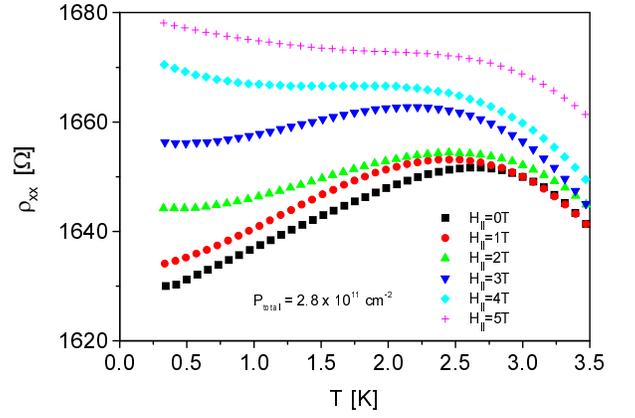}}
\end{picture}
\end{center}
\caption{Low mobility sample. Longitudinal resistance vs. temperature for
different parallel magnetic fields. PMR and suppression of the metallic
characteristics by the parallel field are evident.}
\label{Fig7(low-mu-parallel-magnetic-field)}
\end{figure}

The transition to an insulating behavior in the high mobility sample is
taking place at a very low carrier concentration where the sample becomes
inhomogeneous. We have therefore preferred to study the transition regime
and the insulating phase in a lower mobility sample, $\mu \simeq
20,000$~cm$^{2} /$(V\,s) for $p = 4 \times 10^{11}$~cm$^{-2}$ at $400$~mK,
where the transition occurs at a moderate density.

Following the high mobility data analysis, we study the magnetoresistance
of the low mobility sample and relate it to the temperature dependence of
$\rho_{xx}$ at $H=0$. Since the quantum corrections to the conductivity are
order unity in quantum conductance units, the WL, WAL, and interaction
correction are more pronounced here compared with the high mobility sample.

\begin{figure}
\begin{center}
\setlength{\unitlength}{3.4in}
\begin{picture}(1,0.73)
\put(0.05,0){\includegraphics[width=0.95\unitlength]{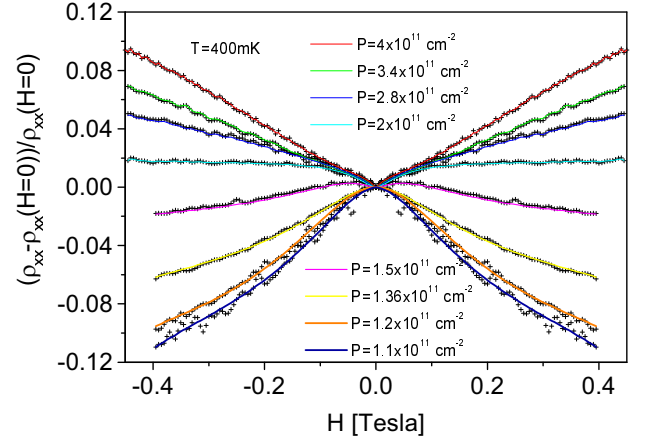}}
\end{picture}
\end{center}
\caption{Low mobility sample. Normalized magnetoresistance at different
carrier densities. Data points are marked with crosses while lines
correspond to the theory discussed in the text. Note the crossover from PMR
to NMR at a density just above the MIT. }
\label{Fig8(low-mu-MR-for-diff.-P)}
\end{figure}
In Fig.~\ref{Fig8(low-mu-MR-for-diff.-P)} we present normalized MR
measurements at $400$~mK for various densities. For high densities, $p \geq
1.5 \times 10^{11}$~cm$^{-2}$, the weak field MR is positive and $d
\rho_{xx} (T, H = 0) / dT > 0$. For $1.5 \times 10^{11}$~cm$^{-2} > p > p_c
= 1.36 \times 10^{11}$~cm$^{-2}$ the MR is negative while $d \rho / d T$ is
still metallic. For $p < p_c$ both the MR and $d \rho_{xx} (T, H = 0) / d
T$ are negative. From the two SdH frequencies one may extract the carrier
densities, $p_l$ and $p_h$, for the lighter and heavier bands respectively,
as depicted in Fig.~\ref{Fig1(populations)}. Below $p_{total} \approx 2
\times 10^{11}$~cm$^{-2}$ the two bands are no longer distinguishable. At
higher densities, the expected Lorentzian PMR is found. The MR curves
remind Bergmann's data \cite{Bergman} on Mg thin films covered with a
fraction of atomic gold layer. In those experiments a crossover from WAL to
WL was found as function of the spin-orbit scattering strength tuned by the
amount of gold. In our experiment, the role of gold is played by the energy
gap between the spin-orbit split bands combined with elastic interband
scattering.

\begin{figure}
\begin{center}
\includegraphics[width=3.4in]{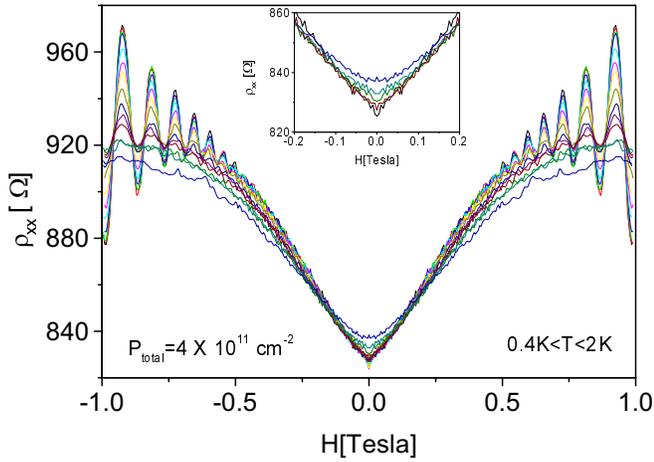}
\end{center}
\caption{Low mobility sample. Magnetoresistance curves at different
temperatures. Inset - magnified view at $T=0.4,0.9,1.3,1.5,2K$. Note the
low $T$, weak field cusp characteristic to WAL.}
\label{Fig9(low-mu-MR-data-for-diff.-T)}
\end{figure}
We focus first on the high density regime. A set of resistance curves vs.\
magnetic field, for $p_{total} = 4 \times 10^{-11}$~cm$^{-2}$, $p_l = 1.7
\times 10^{-11}$~cm$^{-2}$, $p_h = 2.3 \times 10^{-11}$~cm$^{-2}$ and
different temperatures is depicted in
Fig.~\ref{Fig9(low-mu-MR-data-for-diff.-T)}. The PMR is dominated by
classical MR but near the origin, as evident from the inset to
Fig.~\ref{Fig9(low-mu-MR-data-for-diff.-T)}, the experimental points
deviate below the Lorentzian curve and form the well known WAL cusp.
Eq.~(\ref{MR(eq.)}) contains six independent variables; $\rho_\sL (H
\rightarrow \infty)$, $L$, $W$, $F$, $\tau_\varphi$ and $\tau_{so}$. To
find them we take advantage of the different magnetic field scales
characteristic to WL and WAL compared with the classical MR. Using $ H > 10
H_{tr} \sim 500$~Gauss data and ignoring WL or WAL corrections we determine
4 parameters, $\rho_{L} (H \rightarrow \infty)$, $L$, $W$, and $F$. We then
resort to small magnetic fields and find the remaining two parameters,
$\tau_{\varphi}$ and $\tau_{so}$.

\begin{figure}
\begin{center}
\setlength{\unitlength}{3.4in}
\begin{picture}(1,0.72)
\put(0.07,0){\includegraphics[width=0.93\unitlength]{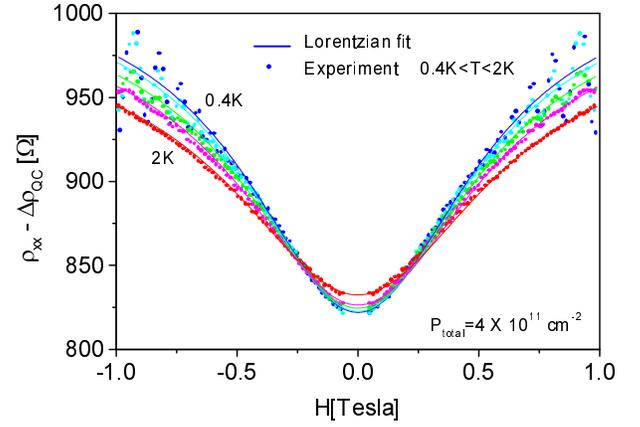}}
\end{picture}
\end{center}
\caption{Low mobility sample. Subtraction of the quantum correction from
the longitudinal resistance according to Eq.~\ref{MR(eq.)} yields the
classical Lorentzian PMR.} \label{Fig10(low-mu-Lorentzian-for-diff.-T)}
\end{figure}
\begin{figure}
\begin{center}
\setlength{\unitlength}{3.4in}
\begin{picture}(1,0.65)
\put(0.05,0){\includegraphics[width=0.95\unitlength]{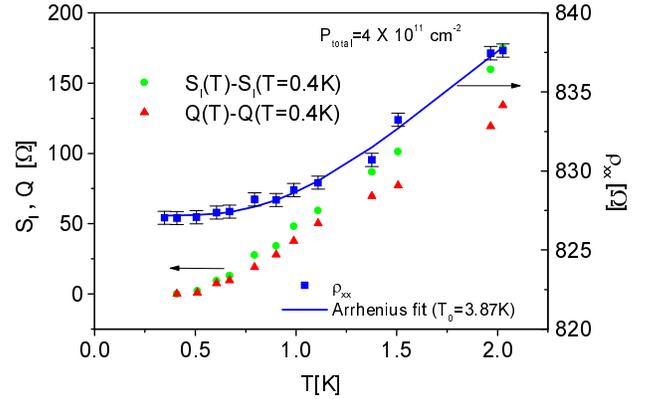}}
\end{picture}
\end{center}
\caption{Low mobility sample. $\Delta S_{l}$ and $\Delta Q$ (left axis) and
the zero field longitudinal resistance (right axis) vs. $T$. Solid line
depicts best fit to Arrhenius dependence upon temperature.}
\label{Fig11(low-mu-rxx,S-vs.-T)}
\end{figure}
Though the procedure for determining the four parameters is similar to the
one used in the high mobility case, the larger quantum corrections lead to
uncertainty in $F$ and consequently, in $S_h$. We have therefore adopted
the following strategy. We first fix $F$ to its theoretical value, $F=0.8$,
and calculate $\rho_\sL (H \rightarrow \infty)$, $L$ and $W$ which in turn
yield $S_l$, $S_h$, and $Q$. However, the resulting $S_h$, and $Q$ display
a slight nonmonotoneity upon temperature below $0.8$~K. Then, for $T <
0.8$~K we tune $F$ to give a monotonous increase of $S_l$, $S_h$, and $Q$
with $T$. The needed change in $F$ is less than $10\%$. Different values of
$F$, ($F = 1, 0.5$) give similar results with different $S_h$. For $F = 1$,
the overall change in $F$ is minimal ($5\%$).
Fig.~\ref{Fig10(low-mu-Lorentzian-for-diff.-T)} depicts the MR data for $F
= 1$ after subtraction of the quantum corrections. The agreement with the
expected Lorentzian is very good. In Fig.~\ref{Fig11(low-mu-rxx,S-vs.-T)},
$S_l - S_l (T = 0.4\text{~K})$, $Q - Q (T = 0.4\text{~K})$, and $\rho_{xx}
(H = 0)$ are depicted vs.\ $T$ for the same total density as in
Fig.~\ref{Fig9(low-mu-MR-data-for-diff.-T)} and $F = 1$. Unlike the high
mobility case, the uncertainty in $F$ and hence, $S_h$, might be
substantial. Consequently we limit ourselves to a qualitative comparison
between $S_l$, $Q$ , and $\rho_{xx} (H = 0)$. Examining
Fig.~\ref{Fig11(low-mu-rxx,S-vs.-T)} one finds a qualitative agreement
between the temperature dependence of the three resistances, in agreement
with the high mobility data.

\begin{figure}
\begin{center}
\includegraphics[width=3.4in]{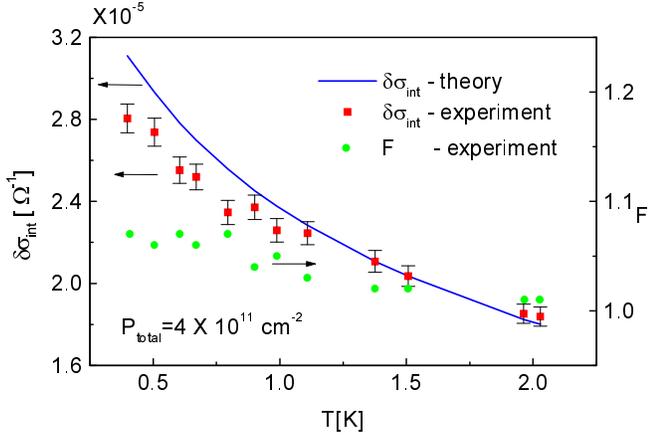}
\end{center}
\caption{Low mobility sample. Left axis: Hole-hole interaction correction
to the conductivity in the diffuson channel, $\protect\delta \protect\sigma
_{int}^{D}$. Right axis: Angular average of the statically screened Coulomb
interaction, $F$. Theoretical value is $F=0.9$.}
\label{Fig12(low-mu-Qcorrection-vs.-T)}
\end{figure}

\begin{figure}
\begin{center}
\setlength{\unitlength}{3.4in}
\begin{picture}(1,0.7)
\put(0.06,0){\includegraphics[width=0.94\unitlength]{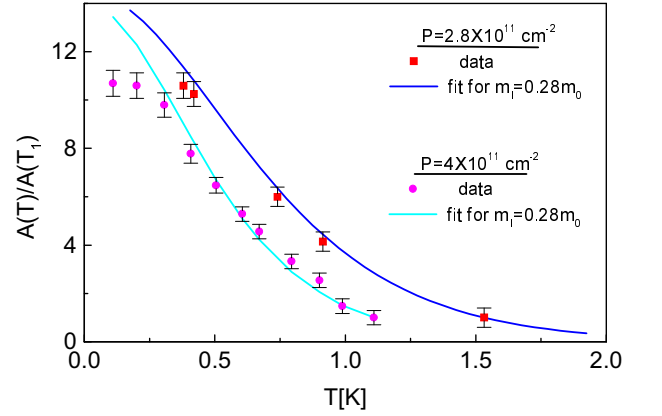}}
\end{picture}
\end{center}
\caption{Shubnikov-de Haas mass analysis. Best fit yields $m_{l}=0.28m_{0}$
where $m_{0}$ is the bare electron mass.}
\label{Fig13(low-mu-mass-from-SdH)}
\end{figure}

Fig.~\ref{Fig12(low-mu-Qcorrection-vs.-T)} compares the theoretical $\delta
\sigma_{int}$ (for $F = 1$), with the measured one as a function of $T$.
For temperatures above $0.8\text{~K}$, the agreement between theory and
experiment is satisfactory, but for temperatures below $0.8\text{~K}$ the
interaction correction is again smaller than the predicted value. We
emphasize that theory was done for two parabolic bands with $m=0.3m_0$. The
pronounced non-parabolicity of the heavy hole band may affect the
interaction contribution significantly. Unlike the high mobility case we do
not observe saturation at low temperatures. From the mass analysis of the
SdH oscillations, depicted in Fig.~\ref{Fig13(low-mu-mass-from-SdH)}, one
finds $m_l = 0.28 m_0$, and no carrier heating above $0.3\text{~K}$.

\begin{figure}
\begin{center}
\setlength{\unitlength}{3.4in}
\begin{picture}(1,0.7)
\put(0.06,0){\includegraphics[width=0.94\unitlength]{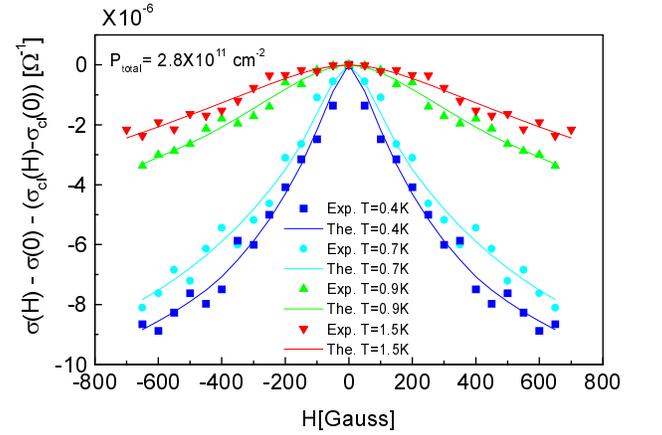}}
\end{picture}
\end{center}
\caption{Low mobility sample. $\protect\sigma (H)-\protect\sigma
(0)-(\protect\sigma _{cl}(H)- \protect\sigma _{cl}(0))$ vs. $T$ for
$p_{total}=2.8\times 10^{11}cm^{-2}$. $\protect\sigma _{cl}(H) $ is the
classical Drude conductivity. Theoretical curves correspond to
eq.~(\ref{WALformula}). Note the PMR in the whole WAL field range.}
\label{Fig14(low-mu-WAL-for-p=2.8)}
\end{figure}
Once $\rho_\sL (H \rightarrow \infty)$, $L$, $W$, and $F$ are determined
one can fit the low magnetic field data to $\sigma (H) - \sigma (H = 0) =
\delta \sigma_\sWL (H) + \delta \sigma_{int} (H) - \delta \sigma_\sWL (0) -
\delta \sigma_{int} (0)$ and extract $\tau_\varphi$ and $\tau_{so}$. Since
the Lorentzian width, $W$, is larger than $H_{tr}$, and since the ratio of
the thermal and dephasing lengths make  the weak field interaction
contribution $\sim 5g^{2}$ times smaller than the WL or WAL one [see
Eqs.~(\ref{WALformula}),~(\ref{Diffusion-correction(eq.)})], the MR for $H
\ll H_{tr}$ is dominated by WL or WAL. The uncertainties in F are hence
reflected in less than $10 \%$ uncertainty in $\tau_\varphi$ and
$\tau_{so}$. Fig.~\ref{Fig14(low-mu-WAL-for-p=2.8)} presents $\sigma (H) -
\sigma (H = 0) - [\sigma_{cl} (H) - \sigma_{cl} (0)]$ for $p_{total} = 2.8
\times 10^{11}$~cm$^{-2}$ where $\sigma_{cl} (H)$ is the classical Drude
part of the conductivity. The data agree well with theory,
Eq.~(\ref{WALformula}). The resulting $\tau_\varphi$ and $\tau_{so}$, for
$p = 2.8 \times 10^{11}$~cm$^{-2}$ are depicted in
Figs.~\ref{Fig15(low-mu-medium-density-a)1/tau-phi-vs.-T,b)tau-so-vs.-T)}
and~\ref{Fig16(low-mu-tau-phi-and-tau-so-vs.-p)}. The error bars reflect
the uncertainty in $F$. The phase breaking rate is linear with temperature,
as expected by theory
(Fig.~\ref{Fig15(low-mu-medium-density-a)1/tau-phi-vs.-T,b)tau-so-vs.-T)}).
The prefactor, however, is five times larger than the theoretical
prediction $1 / \tau_\varphi = T / (\hbar g) \cdot \ln (g/2)$. These
$\tau_\varphi$ values are consistent with those found in n-type Silicon
\cite{Davies} and p-type GaAs \cite{Pepper2} systems. The spin-orbit
scattering time, $\tau_{so}$, is as expected practically independent of
temperature. The extracted times satisfy $\tau_\varphi > \tau_{so}$ and
$\tau_{so}$ comparable to $\tau$. Consequently, the WAL PMR should persist
in the whole range, $0 < H < H_{tr}$. The band splitting, $\epsilon_g$, may
be roughly estimated with the help of the Dyakonov-Perel formula and
compared with the value extracted from the dispersion relation (inset to
Fig.~\ref{Fig5(e-h-continuum)}) deduced from SdH oscillations. Close to the
MIT the Dyakonov-Perel spin precession depends on the relaxation time,
$\tau$, extracted from the decay of the SdH oscillations once the mass has
been determined. Fig.~\ref{Fig13(low-mu-mass-from-SdH)} depicts the SdH
mass analysis for $p = 2.8 \times 10^{11}$~cm$^{-2}$. The envelope analysis
yields $\tau \simeq 10^{-11}$~sec. Substituting $\tau$ in Dyakonov-Perel
expression one finds $\epsilon_g \simeq 0.5$~mev, which agrees with the SdH
value. This result strongly supports the association of the WAL with
scattering between the two spin-orbit split bands and establishes WAL as a
powerful tool for measuring $\epsilon_g$.

\begin{figure}
\begin{center}
\setlength{\unitlength}{3.4in}
\begin{picture}(1,0.67)
\put(0.06,0){\includegraphics[width=0.94\unitlength]{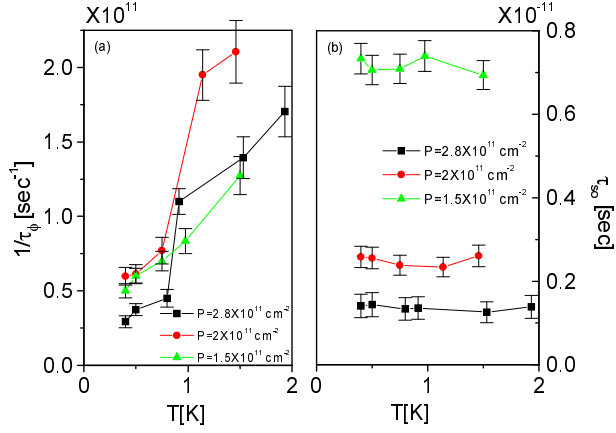}}
\end{picture}
\end{center}
\caption{Low mobility sample. WAL analysis. (a) dephasing rate vs. $T$ at
different densities. (b) spin-orbit scattering time vs. $T$ at different
densities.}
\label{Fig15(low-mu-medium-density-a)1/tau-phi-vs.-T,b)tau-so-vs.-T)}
\end{figure}
We turn now to analyze the data corresponding to densities below $p = 2
\times 10^{11}$~cm$^{-2}$ where the Lorentzian classical MR and the two SdH
frequencies can not be resolved. Then, Eq.~(\ref{MR(eq.)}) takes the single
band form \cite{Choi},
\begin{equation}\label{one-band-MR(eq.)}
\rho_{xx} (H) = \rho_0 - \rho_0^2 \left[ 1 - (\omega_c \tau)^2 \right]
(\delta \sigma_\sWL + \delta \sigma_{int}) \, ,
\end{equation}
where $\rho_0$ is the classical longitudinal resistance. There are 4
parameters to be determined, $\rho_0$, $F$, $\tau_\varphi$ and $\tau_{so}$.
Again, we separate the fitting procedure to $H \gg H_{tr}$ and $H \ll
H_{tr}$. Albeit, this time $H_{tr} > 0.1$~T and consequently the WAL or WL
corrections are important in the whole magnetic field range, $0 < H <
0.7$~T. The four parameters are found iteratively. In the weak field regime
we guess $\rho_0$ and $F$ and fit $\sigma (H) - \sigma (H = 0)$ to theory
with $\tau_\varphi$ and $\tau_{so}$ as two fitting parameters. The
resulting $\tau_\varphi$ and $\tau_{so}$ are plugged into
Eq.~(\ref{one-band-MR(eq.)}) and the high field data are fitted with
$\rho_0$ and $F$ as two adjustable parameters. This procedure converges
after $\sim 10$ iterations to a self consistent solution. The results of
such an analysis for $p = 1.5, 2 \times 10^{11} \text{~cm}^{-2}$ are
presented in
Figs.~\ref{Fig15(low-mu-medium-density-a)1/tau-phi-vs.-T,b)tau-so-vs.-T)}
and~\ref{Fig16(low-mu-tau-phi-and-tau-so-vs.-p)}. Again, $\tau_{so}$ is
practically independent of temperature.
Fig.~\ref{Fig17(low-mu-WAL-for-p=2)} depicts the quantum correction to the
magnetoconductivity $\sigma (H) - \sigma (H = 0)-[\sigma_{cl} (H) -
\sigma_{cl} (0)]$ for $p_{total} = 2 \times 10^{11} \text{~cm}^{-2}$. The
data agree well with Eq.~(\ref{WALformula}). Since $H_{so} < H_{tr}$ the MR
in Fig.~\ref{Fig17(low-mu-WAL-for-p=2)} changes sign from positive to
negative as the field increases. The weak field PMR results from WAL
($\tau_\varphi > \tau_{so}$) and one may use the Dyakonov-Perel formula to
estimate $\epsilon_g$. For $p = 2 \times 10^{11} \text{~cm}^{-2}$ and $p =
1.5 \times 10^{11} \text{~cm}^{-2}$ we find $\epsilon_g \simeq 0.3
\text{~mev} \simeq 0.2 E_\sF$ and $\epsilon_g \simeq 0.2 \text{~mev} \simeq
0.16 E_\sF$, respectively. For $p \geq 2.8 \times 10^{11} \text{~cm}^{-2}$,
the classical PMR as well as the two frequencies of the SdH oscillations
are visible implying $\epsilon_g > \hbar / \tau$. For densities below $2.8
\times 10^{11} \text{~cm}^{-2}$, $\epsilon_g < \hbar / \tau$, and the two
bands can be resolved solely by WAL. These results demonstrate that the
Lorentzian MR as well as the two SdH frequencies disappear when the band
spacing becomes smaller than the level broadening by disorder. The quantum
interference, on the other hand, indicates the existence of two bands down
to lower densities.
\begin{figure}
\begin{center}
\setlength{\unitlength}{3.4in}
\begin{picture}(1,0.65)
\put(0.04,0){\includegraphics[width=0.96\unitlength]{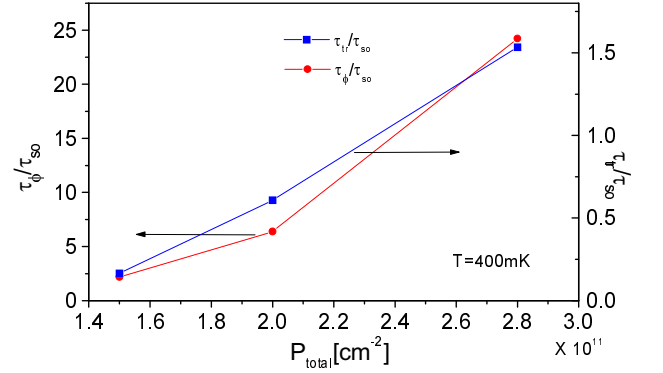}}
\end{picture}
\end{center}
\caption{Low mobility sample. $\protect\tau _{\protect\varphi
}/\protect\tau _{so}$ vs. total carrier density (left axis). $\protect\tau
/ \protect\tau _{so}$ vs. total carrier density (right axis).}
\label{Fig16(low-mu-tau-phi-and-tau-so-vs.-p)}
\end{figure}
\begin{figure}
\begin{center}
\setlength{\unitlength}{3.4in}
\begin{picture}(1,0.72)
\put(0.07,0){\includegraphics[width=0.93\unitlength]{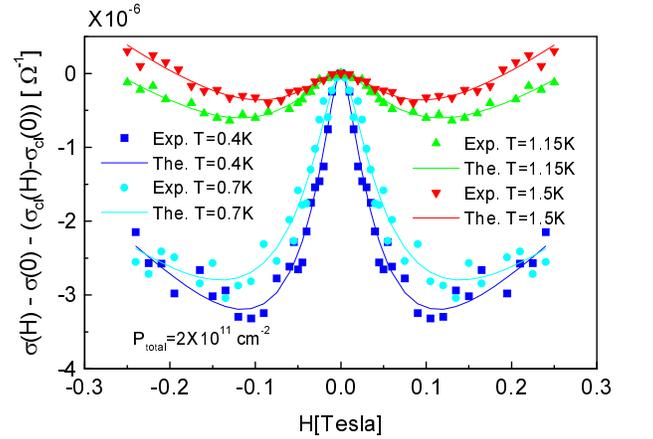}}
\end{picture}
\end{center}
\caption{$\sigma (H) - \sigma (0) - (\sigma_{cl} (H) - \sigma_{cl} (0))$
vs. $T$ for $p_{total} = 2 \times 10^{11}$~cm$^{-2}$. Theoretical curves
correspond to Eq.~(\ref{WALformula}). Note the crossover from PMR to NMR
resulting from the fact that $\tau < \tau_{so}$.}
\label{Fig17(low-mu-WAL-for-p=2)}
\end{figure}

The extracted Coulomb interaction parameter is $F \simeq 1 \pm 0.2$,
practically independent of temperature. This value is in good agreement
with the theoretical prediction, $F = 0.9$. The resulting $\rho_0$
increases with temperature. The origin for this temperature dependence is
not clear. It might result from inelastic inter-band scattering or other
effects such as percolation or temperature dependent screening. Those
mechanisms are discussed in the next section.

\begin{figure}
\begin{center}
\includegraphics[width=3.4in]{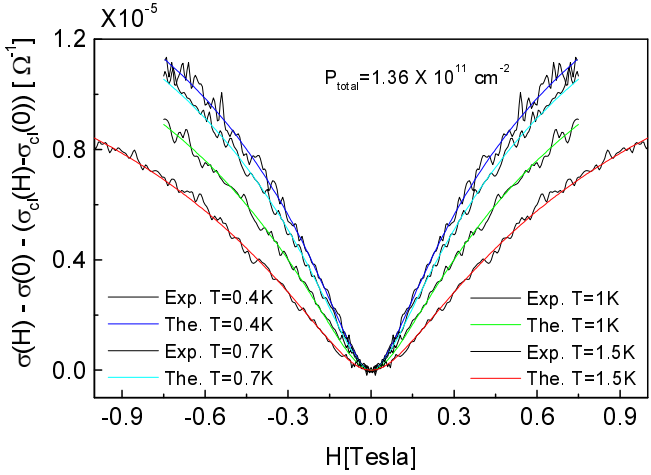}
\end{center}
\caption{Low mobility sample. $\sigma (H)- \sigma (0)-[\sigma_{cl} (H) -
\sigma_{cl} (0)]$ vs. $T$ for $p_{total}=1.36 \times 10^{11}
\text{~cm}^{-2}$. Theoretical curves correspond to the standard WL formula
[Eq.~(\ref{WALformula}) with $\tau_{so} \rightarrow \infty$].}
\label{Fig18(low-mu-WL-for-p=1.36)}
\end{figure}

For $p \leq p_c = 1.36 \times 10^{11} \text{~cm}^{-2},$ WL and interaction
dominate and lead to NMR. At these densities, $g \leq 2.5$, the quantum
corrections are on the order of the classical resistance, and the expansion
in $1 / g$ is only approximate. There are three parameters to be
determined: $F$, $\tau_\varphi$ and $\tau$. Fitting $\sigma (H) - \sigma (H
= 0)$ to theory indicates that WL dominates over interaction. Different
values of $F$ modify the extracted $\tau_\varphi$ by less than $10\%$. One
may therefore set $F=1$, and find $\tau_\varphi$ and $\tau$.
Fig.~\ref{Fig18(low-mu-WL-for-p=1.36)} depicts the data and a best fit for
$p_{total} = 1.35 \times 10^{11} \text{~cm}^{-2}$.
Fig.~\ref{Fig19(low-mu-low-density,1/tau-phi-vs.-T)} displays the extracted
dephasing rate for the three lowest densities. Note that $1 / \tau_\varphi$
extrapolates to a finite value for $T \rightarrow 0$. Similar saturation of
$\tau_\varphi$ has been reported by Brunthaler \emph{et al.}
\cite{Brunthaler} in their Si-MOSFET weak localization measurements. We
emphasize though that we have used WL theory at the border of its validity,
$\delta g \lesssim g$. The extracted $\tau_\varphi$ values might hence be
misleading.
\begin{figure}
\begin{center}
\setlength{\unitlength}{3.4in}
\begin{picture}(1,0.67)
\put(0.06,0){\includegraphics[width=0.94\unitlength]{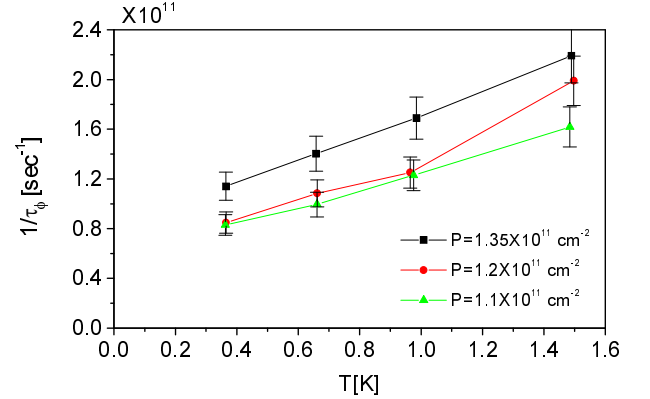}}
\end{picture}
\end{center}
\caption{WL analysis of low mobility sample. Dephasing rate vs. $T$ at
different carrier densities. Note the deviation of $\tau_\varphi^{-1}$ from
linearity in $T$ as discussed in the text.}
\label{Fig19(low-mu-low-density,1/tau-phi-vs.-T)}
\end{figure}

\section{Discussion of data analysis}

\label{Discussion}

At high densities, $p > 2 \times 10^{11} \text{~cm}^{-2}$, we find that the
PMR and temperature dependent resistance are fully consistent with
scattering between the two spin-orbit split bands, WAL, and interaction.
The resistance increase with temperature is mainly a consequence of
inelastic interband scattering. For intermediate densities, $1.36 \times
10^{11} \text{~cm}^{-2} < p < 2 \times 10^{11} \text{~cm}^{-2}$, WAL
indicates two non-degenerate bands. For example, for $p = 2 \times 10^{11}
\text{~cm}^{-2}$ we find $\epsilon_g /E_\sF \simeq 0.2$, implying roughly
$20\%$ difference in densities and resistances of the two bands. For
densities smaller than the critical density, $p_c = 1.36 \times 10^{11}
\text{~cm}^{-2}$, the two bands broadened by disorder merge to form a
single band. The classical PMR as well as the WAL vanish and leave a
conventional, single band WL NMR. Similar MR is also found in other
experiments on p-type GaAs and SiGe \cite{Pepper2,Coleridge2} samples. The
zero field WL crosses over to strong localization as the density is further
reduced. The MIT we observe reflects in our opinion effective cancellation
of the temperature dependence of the resistance due to two competing
processes. On one hand the reminiscent band splitting leads to inelastic
interband scattering and WAL, both characterized by $d \rho_{xx} / d T >
0$. On the other hand, interaction and WL lead to $d \rho_{xx} / d T < 0$.
As the density is reduced, the splitting between the two bands shrinks,
their broadening due to disorder is enhanced and they practically merge to
form a single, doubly degenerate band. In that process, the effect of
inelastic scattering on the resistance is gradually diminished, the WAL
turns into WL ($\tau_\varphi$ becomes comparable to $\tau_{so}$), and
together with the Coulomb interaction, the sample displays an insulating
behavior. The so called MIT is hence just a smooth crossover from two to
one band physics. Note that on the metallic side, near the transition, our
MR data, as well as other data, \cite{Pepper2,Coleridge2} agree well with
WL. Nonetheless, the temperature dependence is opposite to that expected
from WL. There should, hence, be some scattering mechanism with opposite
$T$ dependence compared with WL and very weak dependence upon magnetic
fields. Extrapolating our knowledge from the metallic regime we conjecture
that this mechanism is inelastic interband scattering.

In other experiments on higher mobility 2DHG samples
\cite{Hanein1,Pepper1,Pepper2}, the critical density, $p_c$, turned out to
be lower than that reported here. In Ref.~\onlinecite{Hanein1}, $p_c = 1.25
\times 10^{10} \text{~cm}^{-2}$, and in Refs.~\onlinecite{Pepper1,Pepper2},
$p_c$ is around $5 \times 10^{10} \text{~cm}^{-2}$. We believe that in
those experiments the bands are split down to densities lower than in our
experiment, due to the following reasons: (a) For the mobilities of
Refs.~\onlinecite{Hanein1,Pepper1}, the bands broadening by disorder, and
hence the bands mixing, is small. (b) The quantum well asymmetry and hence
the lack of inversion symmetry are more pronounced in those samples.
Consequently, the band splitting due to spin-orbit interaction is
considerably larger than in our samples even with the lower densities taken
into account. The lack of inversion symmetry is particularly pronounced in
the p-type inverted semiconductor insulator semiconductor (ISIS) structure
\cite{Meirav} used in Ref.~\onlinecite{Hanein1}.

Quantitative estimates of $\epsilon_g$ require detailed band structure
calculations \cite{Goldoni}. Winkler \cite{Winkler} studied band splitting
in 2DHG due to lack of inversion symmetry and was able to show that SdH
analysis underestimates the difference in carrier partition between the two
bands at $H=0$. For a typical confining potential at $p = 2 \times
10^{10} \text{~cm}^{-2}$, Winkler found a $10\%$ difference in hole
densities and $20\%$ difference in their masses. The splitting, $\epsilon_g$,
turned out to be about $0.5 \text{~K}$. These results confirm the role of
two bands in the transport properties of low density 2DHG.

We turn now to discuss some other aspects of the samples that might be
relevant for the MIT. Meir \cite{Meir} pointed out that the 2DHG is
probably inhomogeneous near the MIT and analyzed our data in the context of
classical percolation. He was able to show that the $H = 0$ low temperature
longitudinal resistance can be fitted with $\rho_{xx} = a (p -
p^\star)^{-\gamma}$, with $p^{\star} = 0.9 \pm .05 \times 10^{11}$ and
$\gamma = 1.3 \pm 0.1$. For classical percolation, the critical exponent
$\gamma$ should be $4 / 3 \simeq 1.33$. Identical functional dependence was
found for the data of Hanein \emph{et al.} \cite{Hanein2} Indications of
inhomogeneity are found in thermodynamic measurements as well. Dults and
Jiang measured the compressibility, $\kappa$, of 2DHG \cite{Dultz}, and
found that $d \kappa / d p$ changes sign at the critical density. We also
measured the same quantity in a different way and found similar results
\cite{paper3}. Si and Varma argue that the inverse compressibility should
vanish as the MIT is approached from the metallic side \cite{Varma}. The
argument is that near the transition disorder and Coulomb interaction
create inhomogeneous regions (puddles) which behave like weakly coupled
quantum dots. The breakdown to puddles leads to Coulomb blockades and
incompressibility. Ilani \emph{et al.} \cite{Yacoby} used a single electron
transistor to measure the local compressibility of a 2DHG and found that
below the critical MIT density, the sample is indeed no longer homogenous.
At the moment, the importance of inhomogeneity for the MIT is unclear.
Inhomogeneity should have an interesting effect on the phase breaking
length, $l_\varphi$. When puddles are formed, the diffusion within each
puddle is much faster than that between puddles. In that case one may
imagine a situation where for a limited temperature range $l_\phi$
seemingly saturates to the puddle size.
Fig.~\ref{Fig19(low-mu-low-density,1/tau-phi-vs.-T)} depicts
$\tau_\phi^{-1} \equiv D / l_\phi^2$ (where $l_\phi$ is extracted from WL
and $D$ is the diffusion constant extracted from the global resistance) as
a function of temperature. Indeed $\tau_\phi$ naively extrapolates to a non
zero value for $T \rightarrow 0)$. We do not have though any evidence that
links the deviation of $\tau_\phi^{-1}$ from linear $T$ dependence with
puddle formation.

Another source of a metallic behavior of 2D systems has been proposed by
Gold and Dolgopolov, and Das Sarma and Hwang \cite{Das Sarma}, and was very
recently revisited by Zala \emph{et al.} \cite{Narozhny} These authors
consider impurity screening by the 2DEG. The main dependence of the
dielectric function upon temperature results from $2 k_\sF$ back scattering
of the electrons. The calculation suggests that $\Delta \sigma / \sigma = f
(T / T_\sF)$ where $f$ is some scaling function. The disorder potential in
Si-MOSFET is short ranged and characterized by significant $2k_\sF$
wave-vector components. In modulation doped semiconductors, however, the
separation of the dopant layer from the 2D gas leads to a long range
impurity potential with wave vectors considerably smaller than $k_\sF$.
Nonetheless, Hamilton \emph{et al.} \cite{Hamilton2} took the function $f
(T / T_\sF)$ as universal function and succeeded to collapse their GaAs
hole data (various densities), to a single curve which is linear at small
$T / T_\sF $, and saturates to a constant value at higher values. We tried
a similar analysis but could not find such a universal scaling. Hamilton
\emph{et al.} \cite{Hamilton2} argue that slightly above the critical
density, the asymmetry of the confining potential is irrelevant for the
resistance change with temperature. By fixing $g = k_\sF l$, they claim to
eliminate the influence of the confining potential shape and conclude that
the metallic behavior is due to temperature dependent screening rather than
two bands physics. However, in a recent paper, Papadakis \emph{et al.}
\cite{Papadakis2} show that in less symmetric quantum wells, where subband
splitting is large, the resistance increase with temperature is large and
uncorrelated to the low $T$ resistance. While temperature dependent $2
k_\sF$ screening might be a relevant factor, we find the two band physics
essential for understanding our results.

\section{Theoretical aspects of 2DHG magnetotransport}
\label{Calculation}

\subsection{Arrhenius temperature dependence of the hole-hole
interband scattering rate}
\label{6:sourceSec}

\subsubsection{Qualitative discussion}

In this section we consider the hole-hole scattering contribution to the
resistivity in a two band system. Since the source of band splitting in the
system is spin-orbit coupling, two momenta states in different bands
generally have overlapping spin wave functions. A finite interband
scattering at vanishingly low temperatures (see
fig.~\ref{Fig3(high-mu,rxx,Q,S-vs.-T)}) results from impurity scattering.
At higher temperatures, hole-hole scattering commences and due to the
different mobilities in the two bands, increases the resistance. Similarly
to the $H=0$ Coulomb drag between two layers, one naively expects a $T^{2}$
Coulomb scattering contribution to the resistance. Experimentally we rather
find an unexpected Arrhenius dependence upon temperature. Similar $T$
dependence is found in many experiments probing the metallic phase in 2D.

We turn now to analyze the temperature dependence of hole-hole inter-band
scattering and show that its enhancement by acoustic plasmons leads to an
Arrhenius law for the resistivity at temperatures which are not too low.

As discussed in Section~\ref{6:calcSec} below, the dominant
contribution leading to an Arrhenius $T$ dependence arises from
scattering of a light hole off another hole that changes its band
in the process (light to heavy or vice versa). The contribution of
this process to $S_{\ell }$ is given by
\begin{eqnarray}
S_{\ell }^{(1)} &=&\frac{1}{4\pi ^{2}}\,\frac{h}{e^{2}}\,\frac{1}{Tp_{\ell
}^{2}}\int \frac{d^{2}q}{(2\pi )^{2}}\int_{-\infty }^{\infty }\frac{\hbar
\,d\omega }{\sinh ^{2}(\hbar \omega /2T)}  \label{6:r} \\
&&\;\;\;\;\;\;\;\;\;\; \times \,|V_{{\rm sc} }(q,\omega )|^{2}\,{\rm
Im}\chi _{h\ell }^{k_{x}^{2}}({\bf q},\omega )\,{\rm Im}\chi _{\ell \ell
}(q,\omega )\,,  \nonumber
\end{eqnarray}
with similar expressions for the contributions to $S_{h}$ and $Q$. In
Eq.~(\ref{6:r}), $V_{{\rm sc}}({\bf q},\omega )$ is the screened Coulomb
interaction. The $\chi $ functions are response functions generalized to a
two band system. They are defined explicitly in Eqs.~(\ref {6:chiij})
and~(\ref{6:chigij}). Their imaginary parts, appearing in Eq.~(\ref{6:r}),
are non vanishing whenever it is possible to excite a particle-hole pair
with momentum $\hbar {\bf q}$ and energy $\hbar \omega $ by exciting a
particle from band $i$ to band $j$ (with $i,j$ being $l,l$ or $h,l$).

\begin{figure}
\begin{center}
\includegraphics[width=3in]{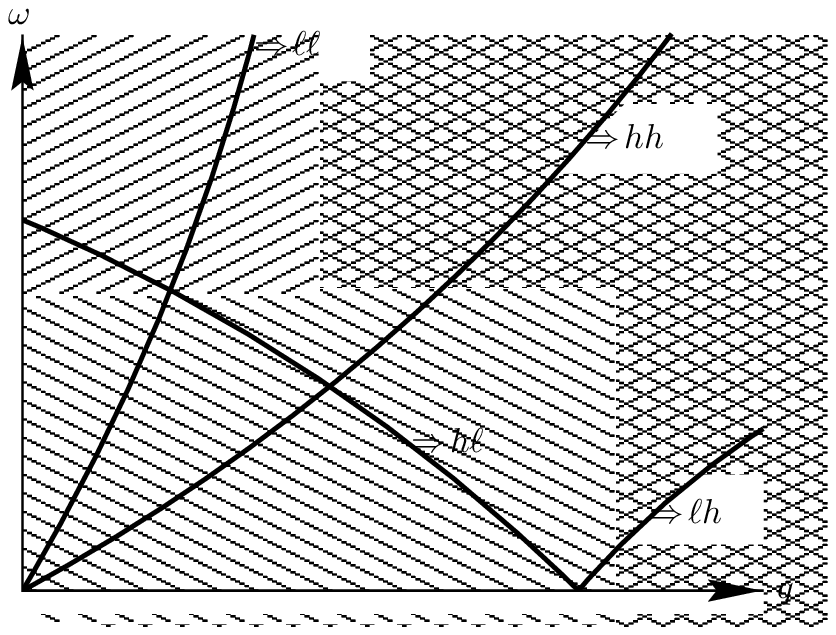}
\end{center}
\caption{The $(q,\protect\omega )$ plane for small $q$ and
$\protect\omega$. The regions in which interband and intraband
particle-hole excitations are possible are marked in the figure. The filled
regions mark the intraband excitations in the light band~($\ell \ell $) and
the interband excitations from the heavy band to the light band ($h\ell $).
The scattering process discussed in the text [Eq.~(\ref{6:r})] is limited
to $q$ and $\protect\omega $ in the overlap of both regions.}
\label{6:qw2figure}
\end{figure}
\begin{figure}
\begin{center}
\includegraphics{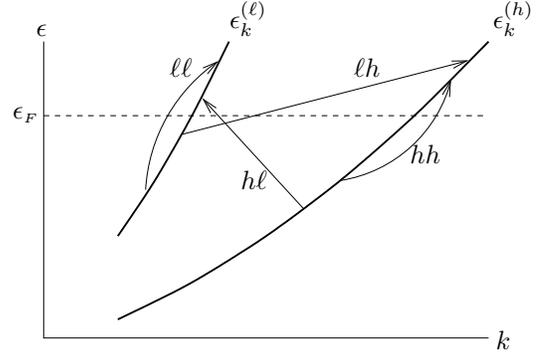}
\end{center}
\caption{The arrows represent the different particle-hole excitations. This
figure is given for illustration only---the real situation is more complex
as momentum space is two-dimensional.}
\label{6:hl2figure}
\end{figure}
\begin{figure}
\begin{center}
\includegraphics[width=3in]{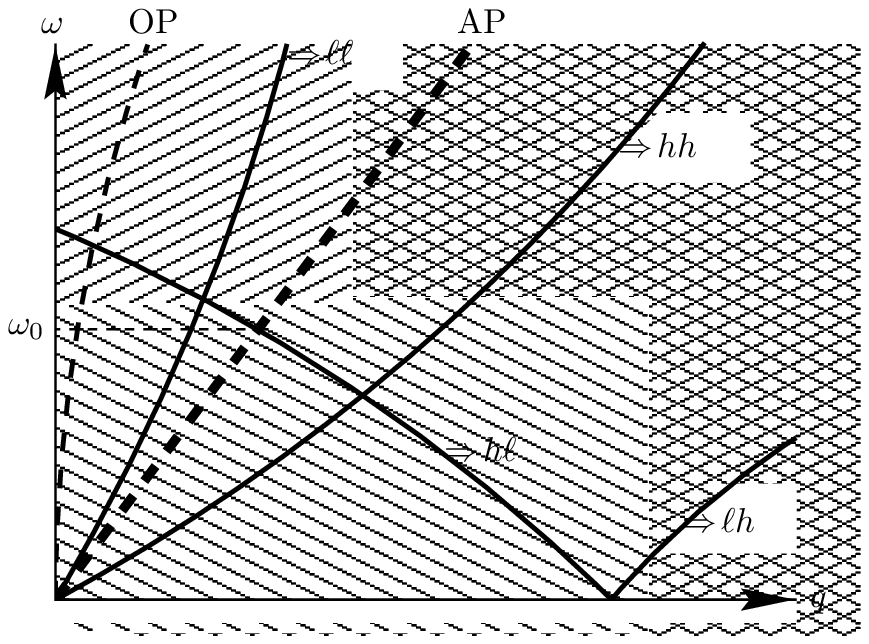}
\end{center}
\caption{The $(q,\protect\omega)$ plane, for small $q$ and
$\protect\omega$. The optical (OP) and acoustic~(AP) plasmon branches are
shown in dashed lines. The regions in which interband and intraband
particle-hole excitations are possible are marked as in
fig.~\ref{6:qw2figure}. The scattering process discussed in the text
[eq.~(\ref{6:r})] is strongly enhanced along the acoustic plasmon line. As
seen in the figure, there is a minimal frequency, $\protect\omega_0$, for
this process.} \label{6:qwfig}
\end{figure}
Fig.~\ref{6:qw2figure} depicts regions in the $(q,\omega )$ plane for which
particle-hole excitations of different types exist (for small $q$
and~$\omega $). Intraband excitations are possible only to the right of a
line whose slope at the origin corresponds to the Fermi velocity in that
band (electron-hole continuum). Interband excitations are forbidden at very
small $q$ and~$\omega$. At $\omega =0$ the minimal wavevector transfer is
$k_\sF^{(h)} - k_\sF^{(\ell )}$, while at $q=0$, the minimal energy for
excitation from the heavy to the light band corresponds to the band
splitting, $\epsilon _{g}$, between the bands at $k_\sF^{(\ell )}$. The
scattering process described by Eq.~(\ref{6:r}) is possible at $q$ and
$\omega $ for which both ${\rm Im}\chi _{\ell \ell }$ and ${\rm Im}\chi
_{h\ell }$ are nonvanishing. The various scattering processes are
illustrated in fig.~\ref{6:hl2figure}.

Due to the existence of two bands, the system is characterized by two
plasmon branches. The optical branch is essentially governed by the motion
of light holes. Its dispersion relation is similar to that of plasmons in a
single-band 2D system, namely, $\omega \propto \sqrt{q}$. The acoustic
branch is characterized by a linear dispersion, $\omega \propto q$,
corresponding to the motion of the heavy holes whose mutual interaction is
screened by the light holes. The acoustic mode is hence similar to a sound
wave in a metal, namely, ion plasma oscillations screened by electrons. The
ions are played here by the heavy holes while the electrons are replaced by
light holes. The acoustic plasmon velocity is approximately given by $v_{p}
= \sqrt{m_{h} / 2m_{\ell }} \, v_\sF^{(h)} $, certifying, $v_\sF^{(h)} <
v_{p} < v_\sF^{(\ell )}$.

The Coulomb scattering depends on the screened Coulomb potential, $V_{{\rm
sc}}(q,\omega )=V_{b}(q)/\varepsilon (q,\omega )$, where $\varepsilon
(q,\omega )$ is the dielectric function of the system. A large contribution
is therefore expected for $q$ and $\omega $ close to the plasmon branches
where $\varepsilon (q,\omega )=0$. The Arrhenius temperature dependence of
the interband scattering rates results from plasmon enhanced scattering. A
similar effect was previously considered for Coulomb drag in double-layer
systems~\cite{FH}.

In fig.~\ref{6:qwfig} we plot the $(q,\omega )$ plane, this time with the
plasmon dispersion lines added. Plasmon enhanced scattering occurs for the
process described above {\em only above a threshold frequency $\omega
_{0}$}. For temperatures $T\ll \hbar \omega _{0}$, the main contribution
comes from frequencies just above the threshold, $\omega _{0}<\omega
<\omega _{0}+T/\hbar $. The contribution of these scattering events is
suppressed by a factor $\propto e^{-\hbar \omega /T}$ due to phase space
considerations but the plasmon enhancement, compensates for that reduction.
The resulting temperature dependence is Arrhenius, $\propto e^{-\hbar
\omega_{0}/T}$.

We turn now to a detailed calculation of the intra and interband scattering
rates resulting from plasmon enhanced hole-hole scattering.

\subsubsection{Detailed calculations - Boltzmann theory for two bands}
\label{6:calcSec}

We start by solving Boltzmann equation for the hole-hole interaction
contribution to the inter and intraband scattering rates.

For a space and time independent distribution function, $f^{(i)}({\bf k})$
($ i=l,h $), the Boltzmann equation in the presence of an electric field
${\bf E}$ takes the form
\begin{equation}
\frac {e \mathbf{E}}{\hbar} \cdot \frac{\partial
f_\mathbf{k}^{(i)}}{\partial \mathbf{k}} = \left( \frac{\partial
f_\mathbf{k}^{(i)}}{\partial t} \right)_\mathrm{coll} \, ,
\label{6:Boltzmanneq}
\end{equation}
The collision integral for hole-hole scattering is given by
\begin{widetext}
\begin{equation}
\left( \frac{\partial f_{{\bf k_{1}}}^{(i)}}{\partial t}\right) _{\! {\rm h
\mbox{-}h}} \!\!\! = \sum_{j i' j'} \sum_{{\bf k}_{2} \mathbf{k}_1' {\bf
k}_{2}^{\prime }} \!\! \left\{ -f_{{\bf k}_{1}}^{(i)}f_{{\bf k}
_{2}}^{(j)}\left( 1-f_{{\bf k}_{1}^{\prime }}^{(i^{\prime })}\right) \left(
1-f_{{\bf k}_{2}^{\prime }}^{(j^{\prime })}\right) W_{{\bf k}_{1}{\bf k}
_{2}\rightarrow {\bf k}_{1}^{\prime }{\bf k}_{2}^{\prime }}^{ij\rightarrow
i^{\prime }j^{\prime }} + f_{{\bf k}_{1}^{\prime }}^{(i^{\prime })}f_{{\bf
k} _{2}^{\prime }}^{(j^{\prime })}\left( 1-f_{{\bf k}_{1}}^{(i)}\right)
\left( 1-f_{{\bf k}_{2}}^{(j)}\right) W_{{\bf k}_{1}^{\prime }{\bf
k}_{2}^{\prime }\rightarrow {\bf k}_{1}{\bf k}_{2}}^{i^{\prime }j^{\prime
}\rightarrow i j}\right\} ,
\end{equation}
where $W_{{\bf k}_{1}{\bf k}_{2}\rightarrow {\bf k}_{1}^{\prime }{\bf k}
_{2}^{\prime }}^{ij\rightarrow i^{\prime }j^{\prime }}$ is the scattering
rate for two holes in states ${\bf k}_{1}$,$i$ and ${\bf k}_{2}$,$j$ to
scatter to ${\bf k}_{1}^{\prime }$,$i^{\prime }$ and ${\bf k}_{2}^{\prime
}$, $j^{\prime }$. Hole-hole scattering satisfies both energy conservation,
$W_{{\bf k}_{1}{\bf k}_{2}\rightarrow {\bf k}_{1}^{\prime }{\bf
k}_{2}^{\prime }}^{ij\rightarrow i^{\prime }j^{\prime }}=w_{{\bf k}_{1}{\bf
k}_{2}\rightarrow {\bf k}_{1}^{\prime }{\bf k}_{2}^{\prime
}}^{ij\rightarrow i^{\prime }j^{\prime }}\,\delta \left( \epsilon
_{k_{1}}^{(i)}+\epsilon _{k_{2}}^{(j)}-\epsilon _{k_{1}^{\prime
}}^{(i^{\prime })}-\epsilon _{k_{2}^{\prime }}^{(j^{\prime })}\right) $,
and momentum conservation, $w_{{\bf k}_{1}{\bf k}_{2}\rightarrow {\bf
k}_{1}^{\prime }{\bf k}_{2}^{\prime }}^{ij\rightarrow i^{\prime }j^{\prime
}}\propto \delta _{{\bf k}_{1}+{\bf k}_{2},{\bf k}_{1}^{\prime }+{\bf
k}_{2}^{\prime }}$. Using the golden rule we have
\begin{equation}\label{6:goldenrule}
w_{\mathbf{k}_1 \mathbf{k}_2 \rightarrow \mathbf{k}_1 - \mathbf{q},
\mathbf{k}_2 + \mathbf{q}}^{i j \rightarrow i' j'} = \frac{2 \pi}{\hbar}
|M_{i i'} (\mathbf{k}_1, \mathbf{k}_1 - \mathbf{q})|^2 \, |M_{j j'}
(\mathbf{k}_2, \mathbf{k}_2 + \mathbf{q} )|^2 \, |V_\mathrm{sc} (q,
\omega)|^2 \, .
\end{equation}
Here, $M_{ij}({\bf k},{\bf k'})$ are the matrix elements between the spin
wavefunctions, derived in Section~\ref{6:modelSec}, below.

For convenience we take $\mathbf{v}_\mathbf{k}^{(i)} = \hbar \mathbf{k} /
m_i$, i.e., neglect non-parabolicity and non-isotropy of the two bands.
Linearizing the collision integral for hole-hole scattering, we have
\begin{eqnarray}\label{6:dfdthh}
\lefteqn{\left( \frac{\partial f^{(i)}({\bf k}_{1})}{\partial t}\right)
_{\rm h\mbox{-}h} \!\! =  -\frac{1}{4T} \sum_{j i' j'} \sum_{\bf k_{2}\bf
q} \int_{-\infty }^{\infty} \frac{\hbar d \omega}{\sinh^2 (\hbar \omega / 2
T)} \, w_{{\bf k}_{1}{\bf k}_{2}\rightarrow {\bf k}_{1}-{\bf q},{\bf k}
_{2}+{\bf q}}^{ij\rightarrow i' j'}\left( \nu _{{\bf k}_{1} - {\bf
q}}^{(i')}+\nu _{{\bf k}_{2}+{\bf q}}^{(j')} - \nu_{{\bf k}_{1}}^{(i)}-\nu
_{{\bf k}_{2}}^{(j)}\right)} \\ & & \;\;\;\;\;\;\; \times \left[
f_{0}\left( \epsilon _{k_{1}}^{(i)}\right) -f_{0} \left( \epsilon
_{k_{1}}^{(i)}-\omega \right) \right] \delta \left( \epsilon _{k_{1}}^{(i)}
- \epsilon_{|{\bf k_{1}}-{\bf q}|}^{(i')} - \hbar \omega \right) \left[
f_{0}\left( \epsilon _{k_{2}}^{(j)}\right) -f_{0}\left( \epsilon
_{k_{2}}^{(j)}+\omega \right) \right] \delta \left( \epsilon
_{k_{2}}^{(j)}-\epsilon_{|{\bf k}_{2}+{\bf q}|}^{(j')} + \hbar \omega
\right) . \nonumber
\end{eqnarray}
We define the intra and interband response functions,
\begin{equation}
\chi _{ij}(q,\omega ) = - \sum_{{\bf k}}\frac{f_{0}\left( \epsilon
_{k}^{(i)}\right) -f_{0}\left( \epsilon _{|{\bf k}+{\bf q}|}^{(j)}\right)
}{ \epsilon _{k}^{(i)}-\epsilon _{|{\bf k}+{\bf q}|}^{(j)} + \hbar \omega
+i\delta } |M_{ij}({\bf k},{\bf k}+{\bf q})|^{2}\,.  \label{6:chiij}
\end{equation}
In addition, we define response functions with momentum (or current)
vertices by adding an appropriate function of ${\bf k}$ to the definition
\begin{equation}\label{6:chigij}
\chi _{ij}^{g({\bf k})}({\bf q},\omega ) = - \sum_{{\bf k}}g({\bf k})\frac{
f_{0}\left( \epsilon _{k}^{(i)}\right) -f_{0}\left( \epsilon _{|{\bf
k}+{\bf q}|}^{(j)}\right) }{\epsilon _{k}^{(i)}-\epsilon _{|{\bf k}+{\bf q}
|}^{(j)} + \hbar \omega + i \delta } |M_{ij}({\bf k},{\bf k}+{\bf q})|^{2}
\, .
\end{equation}
For example, the current-current intraband response function is written in
this notation as $\chi _{ii}^{{\bf jj}}({\bf q},\omega )$, where ${\bf j}=(%
{\bf k}+{\bf q}/2)/m_{i}$. In this example, the function $g({\bf
k})$ depends on ${\bf q}$ as well.

Solving the linearized Boltzman equation we derive the contribution of
hole-hole scattering to the resistivity matrix,
\begin{eqnarray}
\lefteqn{\rho _{ik}^{\rm h\mbox{-}h} = \frac{1}{4 \pi^2} \frac{h}{e^2}
\frac{1}{T p_i p_k} \sum_{j i' j'} \sum_{\bf q} \int_{-\infty }^{\infty}
\frac{\hbar d \omega }{\sinh^2 (\hbar \omega /2T)}
|V_{\rm sc}(q,\omega )|^{2}}  \label{6:rij} \\
&& \times \left\{ {\rm Im}\chi _{i i'}^{k_{x} (k_{x}+q_{x})}(-{\bf q},
-\omega ) {\rm Im} \chi _{j j'} ({\bf q },\omega ) \delta_{i' k} + {\rm Im}
\chi _{i i'}^{k_{x}}(- {\bf q},-\omega ){\rm Im} \chi_{j
j'}^{k_{x}+q_{x}}({\bf q},\omega
)\delta _{j^{\prime }k}\right.  \nonumber \\
&& \qquad \qquad \qquad \left. -\,{\rm Im}\chi_{i i'}^{k_{x}^{2}}(-{\bf q}
,-\omega ){\rm Im} \chi_{j j'}({\bf q},\omega ) \delta_{ik} - {\rm Im}
\chi_{i i'}^{k_{x}} (-{\bf q},-\omega ) {\rm Im} \chi_{j j'}^{k_{x}}({\bf
q},\omega ) \delta_{jk}\right\} \,.  \nonumber
\end{eqnarray}
\end{widetext}

The dominant contribution to Eq.~(\ref{6:rij}) is given by Eq.~(\ref{6:r}).
In this process a hole changes its band by scattering off a hole in the
light band ($i=j=j^{\prime }=\ell $, $i^{\prime }=h$ or $i=i^{\prime
}=j^{\prime }=\ell $, $j=h$ or $i=i^{\prime }=j=\ell $, $j^{\prime }=h$).
Other contributions are readily extracted from Eq.~(\ref{6:rij}). For
example, for the $\ell \ell$ component of the resistivity matrix there are
four additional contributions:

\begin{itemize}

\item  Interband Coulomb drag ($i=i^{\prime }=\ell $, $j=j^{\prime }=h$).
This is the only term remaining in cases of vanishing matrix elements
between the spin wavefunctions of the two bands, e.g., in the absence of
spin-orbit coupling:
\begin{eqnarray}\label{6:drag}
\rho _{\ell \ell }^{{\scriptscriptstyle D}} & = & \frac{1}{8\pi ^{2}} \,
\frac{h}{ e^{2}}\,\frac{1}{Tp_{\ell }^{2}} \int \frac{d^{2}q}{(2\pi )^{2}}
\int_{0}^{\infty }\frac{\hbar \,d\omega }{\sinh ^{2}(\hbar \omega /2T)} \\
&& \;\;\;\;\;\;\;\;\;\;\;\;\; \times \,q^{2}\,|V_{{\rm sc} }(q,\omega
)|^{2}\,{\rm Im}\chi _{\ell \ell }(q,\omega )\,{\rm Im}\chi _{hh}(q,\omega
) \, .  \nonumber
\end{eqnarray}

\item  Band exchange ($i=j'=\ell $, $i^{\prime }=j=h$):
\begin{eqnarray}\label{6:rhoexchange}
\lefteqn{\rho_{\ell \ell }^{(2)} \! = \! \frac{1}{4 \pi^{2}} \,
\frac{h}{e^{2}} \, \frac{1}{T p_{\ell }^{2}} \int \!\!
\frac{d^{2}q}{(2\pi)^{2}} \! \int_{-\infty }^{\infty} \frac{ \hbar \,d
\omega}{\sinh ^{2}(\hbar \omega /2T)}
|V_{{\rm sc}}(q,\omega )|^{2}} \nonumber \\
&& \!\!\!\!\!\!\! \times \, \left\{
 {\rm Im}\chi _{h\ell }^{k_{x}^{2}}({\bf
q},\omega ) {\rm Im} \chi _{h\ell }(q,\omega ) - \left[ {\rm Im}
\chi_{h\ell }^{k_{x}}({\bf q},\omega ) \right]^{2} \right\}
 \, . \;\;\;\;\;\;\;\;\;\;\;\,
\end{eqnarray}

\item Two holes scatter from one band to the other ($i=j=\ell $, $i^{\prime
}=j^{\prime }=h$):
\begin{eqnarray}
\lefteqn{\rho _{\ell \ell }^{(3)} \! = \! \frac{1}{4\pi ^{2}} \,
\frac{h}{e^{2}} \, \frac{1}{Tp_{\ell }^{2}} \int \!\! \frac{d^{2}q}{(2\pi
)^{2}} \! \int_{-\infty }^{\infty } \frac{\hbar \,d\omega }{\sinh
^{2}(\hbar \omega /2T)} |V_{{\rm sc}}(q,\omega )|^{2}} \nonumber \\
&& \!\!\!\!\!\!\! \times \! \left\{ \! {\rm Im}\chi _{\ell h}^{k_{x}^{2}}
\! ({\bf q},\omega ){\rm Im} \chi _{h\ell } (q,\omega ) \! + \! {\rm Im}
\chi _{\ell h}^{k_{x}} \! ({\bf q},\omega ){\rm Im}\chi _{h\ell }^{k_{x} \!
+ q_{x}} \! ({\bf q},\omega ) \! \right\} \! . \nonumber \\
\end{eqnarray}

\item  One particle changes its band by scattering off a heavy hole
($i=\ell $, $j=i^{\prime }=j^{\prime }=h$):
\begin{eqnarray}\label{6:r4}
\rho _{\ell \ell }^{(4)} &=&\frac{1}{4\pi ^{2}}\,\frac{h}{e^{2}}\,\frac{1}{
Tp_{\ell }^{2}}\int \frac{d^{2}q}{(2\pi )^{2}}\int_{-\infty }^{\infty }
\frac{
\hbar \,d\omega }{\sinh ^{2}(\hbar \omega /2T)}  \\
&& \;\;\;\;\;\;\;\;\;\;\;\;\;\;\; \times \,|V_{{\rm sc} }(q,\omega
)|^{2}\,{\rm Im} \chi _{\ell h}^{k_{x}^{2}}({\bf q},\omega )\,{\rm Im}\chi
_{hh}(q,\omega )\,.  \nonumber
\end{eqnarray}
\end{itemize}

The analogous expression for $\rho _{hh}$ is derived directly from
$\rho _{ll}$ by interchanging band indices, $\ell \leftrightarrow
h$. Using momentum conservation, we find that the general
structure of the hole-hole scattering contribution to the
resistivity is given by
\begin{equation}
\rho ^{{\rm h\mbox{-}h}}\propto \left(
\begin{array}{cc}
\displaystyle\frac{1}{n_{\ell }^{2}} & \displaystyle-\frac{1}{n_{\ell }n_{h}}
\\
&  \\
\displaystyle-\frac{1}{n_{\ell }n_{h}} & \displaystyle\frac{1}{n_{h}^{2}}
\end{array}
\right) \,.  \label{6:rhomatrix}
\end{equation}
The latter matrix has a zero determinant and hence, in the absence
of other scattering, allows for dissipationless flow of current of
equal velocity in the two bands.

We now focus on the dominant contribution, Eq.~(\ref{6:r}). We begin with
the calculation of the screened Coulomb interaction. The interaction and
the response functions are described by $2\times 2$ matrices, with indices
corresponding to the two bands. Within the random phase approximation
\begin{equation}
\hat{V}_{{\rm sc}}=\hat{V}_{{\rm bare}}\left( 1+\hat{\chi}\hat{V}_{{\rm
bare} }\right) ^{-1}\,,
\end{equation}
where
\begin{equation}
{\hat{V}}_{{\rm bare}}=\left(
\begin{array}{cc}
V_{b} & V_{b} \\
V_{b} & V_{b}
\end{array}
\right) \,,\;\;\;\hat{\chi}=\left(
\begin{array}{cc}
\chi _{\ell \ell } & \chi _{\ell h} \\
\chi _{h\ell } & \chi _{hh}
\end{array}
\right) \,.
\end{equation}
It follows from these equations that all screened interaction
matrix elements are identical and given by $V_{{\rm sc}}(q,\omega
)=V_{b}(q)/\varepsilon (q,\omega )$, where $V_{b}(q)=2\pi
e^{2}/\epsilon q$ is the bare interaction ($\epsilon $ is the
dielectric constant of the host material). The dielectric function is
given by
\begin{eqnarray}
\lefteqn{\varepsilon (q,\omega ) =} \\ \nonumber & & \!\!\!\! 1 + \left[
\chi _{\ell \ell }(q,\omega )+\chi _{\ell h}(q,\omega )+\chi _{h\ell
}(q,\omega )+\chi _{hh}(q,\omega )\right] V(q) \,. \label{6:plasmoneq}
\end{eqnarray}

Plasmon excitations occur at wavevectors and frequencies which satisfy
$\varepsilon \lbrack q,\omega _{p}(q)-i\gamma _{p}(q)]=0$, where $\omega
_{p}(q)$ is the plasmon frequency and $\gamma _{p}(q)$ its width. To a good
approximation, spin overlap may be ignored in the calculation of plasmons.
Within this approximation, the intraband response functions are given by
\begin{eqnarray}
\lefteqn{\chi _{ii}(q,\omega )=} \\ \nonumber & & \!\!\! \left\{
\begin{array}{ll}
\displaystyle \!\! \frac{m_{i}}{2\pi \hbar ^{2}} \left\{ 1+i\left[ \left(
qv_{{ \scriptscriptstyle F}}^{(i)}/\omega \right) ^{2}-1\right]
^{-1/2}\right\} &
\; {\rm for} \; \omega <qv_{{\scriptscriptstyle F}}^{(i)} \\ &  \\
\displaystyle \!\! \frac{m_{i}}{2\pi \hbar ^{2}} \left\{ 1-\left[ 1-\left(
qv_{{ \scriptscriptstyle F}}^{(i)}/\omega \right) ^{2}\right]
^{-1/2}\right\} & \; {\rm for} \; \omega > q  v_{{\scriptscriptstyle
F}}^{(i)}\,.
\end{array}
\right.  \label{6:intrachi}
\end{eqnarray}
There are two plasmon branches in the system. The optical branch, with
$\omega \propto \sqrt{q}$, results from the solution of $\varepsilon
(q,\omega )=0$ with $\omega >qv_{{\scriptscriptstyle F}}^{(i)}$ in both
bands. The acoustic branch $\omega \propto q$ appears at frequencies,
$\omega <qv_{{\scriptscriptstyle F}}^{(\ell )}$.

Solving for the acoustic branch, we find to lowest order in $\sqrt{m_{\ell
}/m_{h}\text{ }}$that $\varepsilon (q,\omega )=0$ for $\omega
(q)=v_{p}q-i\gamma q$, where
\begin{equation}
v_{p}-i\gamma =\sqrt{\frac{m_{h}}{2m_{\ell }}}v_{{\scriptscriptstyle F}%
}^{(h)}-i\frac{k_{{\scriptscriptstyle F}}^{(h)}}{4k_{{\scriptscriptstyle F}%
}^{(\ell )}}v_{{\scriptscriptstyle F}}^{(h)}
\end{equation}
The screened interaction in the vicinity of the plasmon frequency
is given by
\begin{equation}
V_{{\rm sc}}(q,\omega )=V_{p}\frac{\gamma q}{\omega -v_{p}q+i\gamma q}\,.
\label{6:Vplasmon}
\end{equation}
Here, $V_{p}$ is the interaction strength at the plasmon peak. Expanding the
denominator to second order in $\sqrt{m_{\ell }/m_{h}}$, we find
\begin{equation}
V_{p}=\frac{2\pi \hbar ^{2}}{m_{\ell }}\frac{v_{{\scriptscriptstyle F}%
}^{(\ell )}}{v_{{\scriptscriptstyle F}}^{(h)}}\left( \sqrt{\frac{m_{h}}{%
2m_{\ell }}}+i\frac{k_{{\scriptscriptstyle F}}^{(h)}}{k_{{\scriptscriptstyle %
F}}^{(\ell )}}\right) ^{-1}\,.
\end{equation}
In the equations above we have used $\sqrt{m_{\ell }/m_{h}}$ as a
small parameter. At low densities this approximation breaks down
but the plasmon dispersion remains linear and the interaction can
still be approximated in the vicinity of the plasmon pole by
eq.~(\ref{6:Vplasmon}) with slightly different values of $v_{p}$,
$\gamma $, and $V_{p}$. The integral over frequency in (\ref{6:r})
is approximated in the vicinity of the plasmon frequency using
\begin{equation}
|V_{{\rm sc}}(q,\omega )|^{2}=\pi \gamma q|V_{p}|^{2}\delta (\omega
-v_{p}q)\,.  \label{6:Vabs2}
\end{equation}

\begin{figure}
\begin{center}
\includegraphics{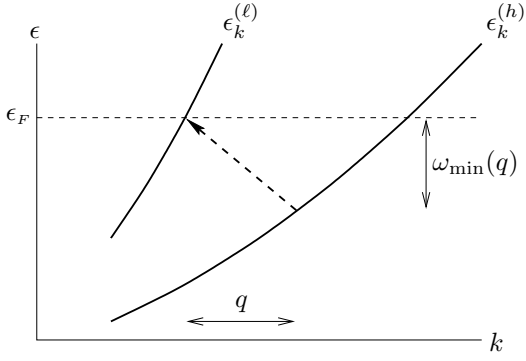}
\end{center}
\caption{Dashed arrow represents particle-hole excitation from the heavy to
the light band at the minimal energy $\protect\omega_{min}(q)$ for a given
momentum $q$ (for $q < k_\sF^{(h)} - k_\sF^{(\ell)}$). This transition
corresponds to excitation of a heavy hole at $k_\sF^{(\ell)} + q$ to the
light band at $k_\sF^{(\ell)}$ in exactly the same direction in momentum
space.} \label{6:hlfigure}
\end{figure}
We turn now to the calculation of the interband response function. An
interband particle-hole excitation from the heavy to the light band at
wavevector ${\bf q}$ and frequency $\omega $ may occur at wavevectors ${\bf
k}$ satisfying $\hbar \omega =\epsilon _{k}^{(\ell )}-\epsilon _{|{\bf k} -
{\bf q}|}^{(h)}$. Possible ${\bf k}$ values are dictated by the $\delta $
function in the expression for ${\rm Im}\chi _{h\ell }$,
\begin{eqnarray}\label{6:interchi}
\lefteqn{\rm Im\chi _{h\ell }^{k_{x}^{2}}(q,\omega ) = \pi \sum_{{\rm
k}}(k_{x}-q_{x})^{2} |M_{h\ell }({\bf k}-{\bf q},{\bf k} )|^{2} } \\
&&  \;\;\;\;\; \times \left[ f_{0}\left( \epsilon _{|{\bf k}-{\bf
q}|}^{(h)}\right) -f_{0}\left( \epsilon _{k}^{(\ell )}\right) \right] \,
\delta \left( \epsilon _{|{\bf k}-{\bf q}|}^{(h)}-\epsilon _{k}^{(\ell
)}+\hbar \omega \right) . \nonumber
\end{eqnarray}
The possible phase space for interband particle-hole excitations
is limited by a minimal frequency, $\omega _{{\rm min}}(q)$. This
 frequency is illustrated in fig.~\ref{6:qw2figure} as the
threshold for particle-hole excitation from the heavy to the light
band. Such excitation is realized when the initial heavy hole
momentum, ${\bf k}-{\bf q}$, and the final light hole momentum,
${\bf k}$, are parallel and the final energy in the light band is
equal to the Fermi energy (fig.~\ref{6:hlfigure}). The angle,
$\theta $, between ${\bf k}$ and ${\bf q}$ for such an excitation
is $\pi $. We are interested in excitations with frequencies close
to $\omega _{{\rm min}}(q)$. Therefore, the angle $\theta $ is
close to $\pi $, and $\lambda =1+\cos \theta $ is a small
parameter.

Using (\ref{6:overlap}), the interband spinor overlap appearing in (\ref
{6:interchi}) is given to leading order in $\lambda $ (and approximating $k
= k_\sF^{(\ell )}$) by
\begin{equation}
|M_{h\ell }({\bf k}-{\bf q},{\bf k})|^{2}=\frac{9}{2}{\cal
C}_{1}\frac{q^{2} }{\left( k_{{\scriptscriptstyle F}}^{(\ell )}+q\right)
^{2}}\lambda \,,
\end{equation}
where ${\cal C}_{1}=\left( \langle \xi_\sH^{(h)}|\xi_\sH^{(\ell )}\rangle
+1/3\langle \xi_\sL^{(h)}|\xi_\sL^{(\ell )}\rangle \right) ^{2}$ measures
heavy/light hole mixing and its magnitude is approximately 1 (exactly 1 in
the absence of mixing). The integrals in (\ref {6:interchi}) may now be
calculated to leading order in $\lambda $ (using $ d\theta =d\lambda
/\sqrt{2\lambda }$) to yield
\begin{eqnarray}
{\rm Im}\chi _{h\ell }(q,\omega ) &=&\frac{3\,{\cal C}_{1}}{2\sqrt{2}\,\pi
} \frac{k_{{\scriptscriptstyle F}}^{(\ell )}q^{2}}{\hbar ^{2}\left( k_{{
\scriptscriptstyle F}}^{(\ell )}+q\right) ^{2}\Delta v(q)}\lambda _{{\rm
max}
}^{3/2}\,  \label{6:chihl} \\
\Delta v(q) &=&v_{{\scriptscriptstyle F}}^{(\ell )}-v_{k_{{
\scriptscriptstyle F}}^{(\ell )}+q}^{(h)}\text{ },
\end{eqnarray}
and
\begin{equation}
\lambda _{{\rm max}}=\frac{m_{h}}{\hbar k_{{\scriptscriptstyle F}}^{(\ell
)}q }\left[ \omega -\omega _{{\rm min}}(q)\right] .
\end{equation}
$\lambda _{{\rm max}}$ represents the maximal deviation of the
particle-hole excitation from $\theta =\pi $ with $k=k_{{
\scriptscriptstyle F}}^{\ell }$. Note that ${\rm Im}\chi _{h\ell }(q,\omega
)\propto \left[ \omega -\omega _{{\rm min}}(q)\right] ^{3/2}$. For the
calculation of the resistivity $\rho _{ij}^{(1)}$ we need ${\rm Im}\chi
_{h\ell }^{k_{x}^{2}}$, eq.~(\ref{6:interchi}). Since the main contribution
comes from a small range of ${\bf k}$ around $-k_{{\scriptscriptstyle F}
}^{(\ell )}\hat{{\bf q}}$, we readily obtain
\begin{equation}
{\rm Im}\chi _{h\ell }^{k_{x}^{2}}({\bf q},\omega )=\left( k_{{
\scriptscriptstyle F}}^{(\ell )}+q\right) ^{2}\cos ^{2}\varphi \,{\rm Im}
\chi _{h\ell }(q,\omega )\,,  \label{6:chikx2hl}
\end{equation}
where $\varphi $ is the angle between ${\bf q}$ and $\hat{{\bf x}}$.

We now have all the ingredients needed for evaluating $S_{l}$,
Eq.~(\ref{6:r}). The interaction term, $|V_{{\rm sc}}|^{2}$, is given by
Eq.~(\ref{6:Vabs2} ), so that all terms are calculated along the acoustic
plasmon dispersion line, $\omega =v_{p}q$. The interband response function
${\rm Im}\chi _{h\ell }^{k_{x}^{2}}$ is given by
Eqs.~(\ref{6:chihl})--(\ref{6:chikx2hl}), limiting the integration range to
$q > q_{0}$ (or $\omega > \omega _{0} = v_{p} q_{0}$). The intraband
response of the light holes is given by
\begin{equation}
{\rm Im}\chi _{\ell \ell }(q,\omega )=\frac{m_{\ell }}{2\pi \hbar ^{2}}\frac{%
\omega }{qv_{{\scriptscriptstyle F}}^{(\ell )}}\frac{|M_{\ell \ell }|^{2}}{%
\sqrt{1-\left( \omega /qv_{{\scriptscriptstyle F}}^{(\ell )}\right) ^{2}}}\,.
\end{equation}
Here, $|M_{\ell \ell }|^{2}=|M_{\ell \ell }({\bf k},{\bf k}+{\bf q})|^{2}$ (%
\ref{6:overlap}), calculated at a wavevector ${\bf k}$ contributing to ${\rm %
Im}\chi _{\ell \ell }(q,\omega )$ (since $\hbar \omega \ll E_{{%
\scriptscriptstyle F}}$, the matrix elements are approximately the
same). The integral (\ref{6:r}) is calculated assuming $%
T<\hbar \omega _{0}$, hence, $\sinh ^{-2}(\hbar \omega /2T)\approx e^{-\hbar
\omega /T}/4$. The integral is thus limited to the vicinity of $q=q_{0}$ and
$\omega =\omega _{0}$. The temperature dependence of $S_{l}^{(1)}$ is
readily evaluated,
\begin{eqnarray}
S_{l}^{(1)} & \propto & \frac{1}{T}e^{-\hbar \omega
_{0}/T}\int_{q_{0}}^{\infty }dq\,(q-q_{0})^{3/2}e^{-\hbar v_{p}(q-q_{0})/T}
\nonumber \\ & \propto & T^{3/2}e^{-\hbar \omega _{0}/T}\,.
\end{eqnarray}
We thus find that as a result of the minimal frequency for interband
excitations, $\rho _{ij}^{(1)}$ displays Arrhenius dependence upon $T$ with
a characteristic energy $T_{0}=\hbar \omega _{0}$. The magnitude and the $%
T^{3/2}$-dependence of the prefactor are determined by details in
the vicinity of the threshold for excitations. Collecting all
 terms and converting the scattering rate into resistivity one
obtains
\begin{eqnarray}
\lefteqn{ \rho _{ij}^{(1)} = (-1)^{\delta _{ij}} \frac{9\sqrt{\pi }}{8}
{\cal C}} \\ \nonumber & & \times \, \frac{h}{ e^{2}}\frac{n_{\ell
}^{2}}{n_{i}n_{j}}\left( \frac{m_{h}}{m_{\ell }}\right) ^{3/2}\left(
\frac{q_{0}}{k_{{\scriptscriptstyle F}}^{(\ell )}}\right) ^{5/2}\left(
\frac{T}{E_{{\scriptscriptstyle F}}}\right) ^{3/2}e^{-\hbar \omega
_{0}/T}\,.  \label{6:rho1result}
\end{eqnarray}
The term ${\cal C}$ is given by
\begin{eqnarray}
\lefteqn{ {\cal C}^{-1} = \frac{1}{{\cal C}_{1}|M_{\ell \ell }|^{2}}} \\
\nonumber & & \! \times \, \left| 1+i\sqrt{ \frac{2m_{\ell
}}{m_{h}}}\frac{k_{{\scriptscriptstyle F}}^{(h)}}{k_{{ \scriptscriptstyle
F}}^{(\ell )}}\right| ^{2}\left( 1-\frac{v_{k_{{ \scriptscriptstyle
F}}^{(\ell )}+q_{0}}^{(h)}}{v_{{\scriptscriptstyle F} }^{(\ell )}}\right)
\sqrt{1-\left( \frac{v_{p}}{v_{{\scriptscriptstyle F} }^{(\ell )}}\right)
^{2}} \, .
\end{eqnarray}

We have considered here the plasmon enhanced contribution to one of the
scattering processes described in Section~\ref{6:calcSec}. With the help of
Fig.~\ref{6:qwfig} we note that the only other plasmon enhanced scattering
process involves carrier exchange between bands as described by
$\rho_{ij}^{(2)}$, Eq.~(\ref{6:rhoexchange}). The latter contribution
comprises, however, two terms that cancel each other within our
approximation for the interband response function. Another possibility
involving thermal activation of ${\rm Im}\chi _{ij}$ beyond its zero
temperature boundaries, in the Coulomb drag term, is discussed in
Ref.~\onlinecite{Sivan}. Thermal activation of plasmons leads to an
Arrhenius dependence as well but with a $T^{7/2}$ prefactor. \cite{Yaib}
That contribution is hence smaller at low temperatures. A full quantitative
calculation takes all contributions into account.

A $T^{2}$ contribution to the resistivity should arise from the different
scattering processes at frequencies $\omega <T$. Preliminary calculations
indicate this term should be observable in the experiment. Its absence is
hence surprising. This discrepancy between theory and experiment might be
related to one or more of the approximations involved in the calculation.
These include the use of response functions calculated for $q\ll
k_{{\scriptscriptstyle F}}$ at large wavevectors, $q\approx
2k_{{\scriptscriptstyle F}}$, and the neglect of the anisotropy and
nonparabolicity of the heavy band.

\subsection{Quantum interference corrections}

\label{6:QuantumSec}

In this section we discuss quantum interference corrections to the
conductivity. We assume a short-range interaction and take the overlap
between spin wavefunctions of the two bands, Eq.~(\ref{6:overlap}) into
account. We find that in the limit of a small interband gap, the physics is
similar to that of electrons in the presence of spin-orbit interaction.
Quantum interference corrections in this case were previously studied
theoretically in refs.~\cite{HLN80,RB85,GDK98b,Lyanda-Geller,Averkiev1}.
Having calculated the Cooperon, the usual derivation (see, e.g,
ref.~\cite{HLN80}) is followed to obtain the correction to the
conductivity.

\begin{figure}
\begin{center}
\includegraphics[width=3.4in]{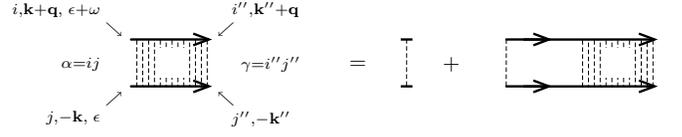}
\end{center}
\caption{Diagram for the Cooperon (see Eq.~\ref{6:Cooperoneq}) The dashed
line describes the short-range disorder.} \label{6:cooperonfig}
\end{figure}
The following notation is used: Band indices of one particle are marked by
Roman letter ($i,j,\ldots $) that take the value $\ell $ or $h$. Greek
letters ($\alpha ,\beta ,\ldots $) denote band indices of the Cooperon two
particles, taking the values $\ell \ell $, $\ell h$, $h\ell $ or $hh$. In
this notation, the Cooperon is a $4\times 4$ matrix with Greek indices
describing the bands of the incoming and outgoing particles. The Cooperon
is given diagrammatically in fig.~\ref{6:cooperonfig}. At a total momentum
${\bf q}=0$ and a total energy $\omega =0$ it is given by (in this section
we use units in which $\hbar = 1$)
\begin{eqnarray}\label{6:Cooperoneq}
\lefteqn{ C_{\alpha \gamma }({\bf k},{\bf k^{\prime \prime }}) = {\cal
V}_{\alpha \gamma }({\bf k},{\bf k^{\prime \prime }}) } \\
\nonumber & & \!\!\!\! + \, \sum_{\beta }\int \!\frac{d^{2}k^{\prime
}}{(2\pi )^{2}}{\cal V}_{\alpha \beta }({\bf k},{\bf k^{\prime }}
)G_{i^{\prime }}^{+}({\bf k},\epsilon )G_{j^{\prime }}^{-}(-{\bf
k},\epsilon )C_{\beta \gamma }({\bf k^{\prime }},{\bf k^{\prime \prime }})
,
\end{eqnarray}
The Green function is given by
\begin{equation}
G_{j}^{\pm }({\bf k},\epsilon )=\frac{1}{\epsilon -\epsilon _{{\bf k}%
}^{(j)}\pm i/2\tau _{j}}\,,
\end{equation}
where $\tau _{j}$ is the mean free time in the $j^{{\rm th}}$ band,
calculated from the disorder within the Born approximation. The Green
functions in Eq.~(\ref{6:Cooperoneq}) limit the contributing momenta to the
vicinity of the Fermi surface. Taking the momenta on the Fermi surface, the
integration over the magnitude ${\bf k^{\prime }}$ is performed to give
\begin{equation}\label{6:Cooperoneq2}
C_{\alpha \gamma }(\theta \! , \theta ^{\prime \prime })={\cal V}_{\alpha
\gamma }(\theta \! , \theta ^{\prime \prime })+\sum_{\beta } \!\! \int \!
\frac{d\theta ^{\prime }}{2\pi }{\cal V}_{\alpha \beta }(\theta \! , \theta
^{\prime })\Pi _{\beta }C_{\beta \gamma }(\theta ^{\prime } \! , \theta
^{\prime \prime }) .
\end{equation}
with
\begin{equation}
\Pi _{\alpha =ij}=\int \frac{d^{2}k}{(2\pi )^{2}}G_{i}^{(+)}({\bf k}%
,\epsilon )G_{j}^{(-)}(-{\bf k},\epsilon )\,,
\end{equation}
Finally, the disorder line is given by
\begin{equation}
{\cal V}_{\alpha =ij,\beta =i^{\prime }j^{\prime }}(\theta ,\theta ^{\prime
})=\left\langle V^{2}\right\rangle M_{ii^{\prime }}(\Delta \theta
)M_{jj^{\prime }}(\Delta \theta )\,.  \label{6:disorder}
\end{equation}
with $\left\langle V^{2}\right\rangle$ being the disorder
correlation function. Note the overlap between spin wavefunctions
(\ref{6:overlap}) is taken here on the Fermi surface and therefore
depends on the angle difference $\Delta \theta =\theta ^{\prime
}-\theta $ only.

We discuss below two limits; A large band gap, $\epsilon _{g}\tau \gg 1$,
as expected at high densities, and a small band gap, $\epsilon _{g}\tau \ll
1$, which becomes relevant as the density approaches the metal-insulator
transition. In both cases, for a short range disorder, the Cooperon
equation can be solved exactly by calculating $C=(1-{\cal V}\Pi )^{-1}{\cal
V}$.

We start with the high density limit, $\epsilon _{g}\tau \gg 1$.
In this case, the bare particle-particle propagator $\Pi _{\alpha
}$ at ${\bf q}=0$, $\omega =0$ is small for two particles in
different bands. Neglecting this term, we find
\begin{equation}
\Pi _{\alpha =ij} = \! \int \!\! \frac{d^{2}k}{(2\pi )^{2}}G_{i}^{(+)}({\bf
k} ,\epsilon )G_{j}^{(-)}(-{\bf k},\epsilon )=\left\{
\begin{array}{ll}
m_{i}\tau _{i} & i=j \\
0 & i\neq j \, .
\end{array}
\right.
\end{equation}
The calculation reduces to $2\times 2$ matrices with $\ell \ell $ and $hh$
indices only. The mean free time is found in the Born approximation, using
Eq.~(\ref{6:overlap}), to be
\begin{eqnarray}
\frac{1}{\tau _{i}} &=&2i\sum_{i^{\prime }}\int \frac{d^{2}k^{\prime }}{
(2\pi )^{2}}G_{i^{\prime }}^{+}({\bf k^{\prime }},\epsilon )\left\langle
V^{2}\right\rangle M_{ii^{\prime }}(\Delta \theta )M_{i^{\prime }i}(-\Delta
\theta )  \label{6:taui} \nonumber \\
& = & \frac{\left\langle V^{2}\right\rangle }{2} \left\{ m_{i}\left[ \left(
\xi _{{\scriptscriptstyle H}}^{(i)}\right) ^{4}+\left( \xi _{{
\scriptscriptstyle L}}^{(i)}\right) ^{4}\right] \right. \\ & & \left.
\left. + \, m_{j}\left[ \left( \xi _{{ \scriptscriptstyle H}}^{(i)}\xi
_{{\scriptscriptstyle H}}^{(j)}\right) ^{2}+\left( \xi
_{{\scriptscriptstyle L}}^{(i)}\xi _{{\scriptscriptstyle L} }^{(j)}\right)
^{2}\right] \right\} \right| _{j\neq i}\, . \nonumber
\end{eqnarray}

The following observations simplify the calculation: In the equation for
the Cooperon, (\ref{6:Cooperoneq2}), the disorder line depends on $\Delta
\theta =\theta ^{\prime }-\theta $, and $\Pi _{\beta }$ is
angle-independent. It is therefore possible to Fourier transform the
equation to angular momentum space. Using Eqs.~(\ref{6:overlap}) and
(\ref{6:disorder}), we note the only nonvanishing terms in ${\cal
V}_{\alpha \beta }(\Delta \theta )=\sum_{m}{\cal V}_{\alpha \beta
}^{(m)}e^{im\Delta \theta }$ have $0\leq m\leq 6$.

The Cooperon equation (\ref{6:Cooperoneq2}) takes the form,
\begin{equation}
C_{\alpha \gamma }^{(m)}={\cal V}^{(m)}+\sum_{\beta }{\cal V}_{\alpha \beta
}^{(m)}\Pi _{\beta }C_{\beta \gamma }^{(m)}\,,
\end{equation}
where $C_{\alpha \beta }(\theta ,\theta ^{\prime
})=\sum_{m}C_{\alpha \beta }^{(m)}e^{im(\theta ^{\prime }-\theta
)}$. The Cooperon has a large contribution whenever the matrix
${\cal V}_{\alpha \beta }^{(m)}\Pi _{\beta } $ has an eigenvalue
close to 1. This happens for $m=3$ only, where
\begin{eqnarray}
\lefteqn{{\cal V}^{(3)} =  \frac{\left\langle V^{2}\right\rangle }{2} } \\
\nonumber & &  \!\!\!\!\!\!\! \times \! \left( \!
\begin{array}{cc}
\left( \xi _{{\scriptscriptstyle H}}^{(\ell )}\right) ^{4} \! + \! \left(
\xi _{{ \scriptscriptstyle L}}^{(\ell )}\right) ^{4} & \left( \xi _{{
\scriptscriptstyle H}}^{(\ell )}\xi _{{\scriptscriptstyle H}}^{(h)}\right)
^{2} \! + \! \left( \xi _{{\scriptscriptstyle L}}^{(\ell )}\xi
_{{\scriptscriptstyle
L}}^{(h)}\right) ^{2} \\
&  \\
\left( \xi _{{\scriptscriptstyle H}}^{(h)}\xi _{{\scriptscriptstyle H}
}^{(\ell )}\right) ^{2} \! + \! \left( \xi _{{\scriptscriptstyle
L}}^{(h)}\xi _{{ \scriptscriptstyle L}}^{(\ell )}\right) ^{2} & \left( \xi
_{{ \scriptscriptstyle H}}^{(h)}\right) ^{4} \! + \! \left( \xi
_{{\scriptscriptstyle L} }^{(h)}\right) ^{4}
\end{array}
\! \right) \! .
\end{eqnarray}
Since $\Pi _{\alpha }=0$ for $\alpha =\ell h$ or $h\ell $, we write ${\cal
V} ^{(3)}$ as a $2\times 2$ matrix with $\ell \ell $ and $hh$ indices
 only. Using (\ref{6:taui}), the matrix ${\cal V}_{\alpha \beta
}^{(3)}\Pi _{\beta }$ has an eigenvalue 1, with eigenvector
$\propto (-\tau _{2},\tau _{1})$.

We show now that quantum interference corrections lead to weak
antilocalization. Using the eigenvector of ${\cal V}_{\alpha \beta
}^{(3)}\Pi _{\beta }$, the Cooperon takes the form
\begin{equation}
C(\Delta \theta )=\left(
\begin{array}{cc}
\tau _{2}/\tau _{1} & -1 \\
-1 & \tau _{1}/\tau _{2}
\end{array}
\right) \frac{\left\langle V^{2}\right\rangle }{(Dq^{2}-i\omega )\tau }
e^{i3\Delta \theta }\,.
\end{equation}
The diffusion constant $D$, and the mean free time $\tau $, reflect the
properties of both bands. In a weak localization diagram, each of the
particle lines at one end of the Cooperon is connected with the {\em other}
particle line at the other end of the Cooperon. Because the momenta of the
two lines is opposite, $\Delta \theta =\pi $ in the expression for the
Cooperon and $e^{i3\Delta \theta }=-1$. This $-1$ term, which does not
exist for the Cooperon in the single-band case, changes the overall sign of
the diagram and leads to weak antilocalization.

In the limit of a large band gap we thus find weak antilocalization. We
turn now to the opposite limit; a small gap, $\epsilon _{g}\tau <1$. In
that case we neglect differences in the density of states and scattering
times in the two bands, as well as the bulk light hole contribution to the
spin overlap functions, $\xi _{{\scriptscriptstyle L}}^{(i)}$. The
calculation is strictly a leading order expansion in $k$. Here, the
Cooperon equation (\ref {6:Cooperoneq}) is not reduced to a $2\times 2$
matrix since the propagation of the two Cooperon particles in two different
bands should be taken into account. The bare particle-particle propagator
at ${\bf q}=\omega =0$ takes the form
\begin{equation}
\Pi =m\tau \left(
\begin{array}{cccc}
1 & 0 & 0 & 0 \\
0 & \frac{1}{1+i\epsilon _{g}\tau } & 0 & 0 \\
0 & 0 & \frac{1}{1-i\epsilon _{g}\tau } & 0 \\
0 & 0 & 0 & 1
\end{array}
\right) \,,
\end{equation}
where indices are taken in the order $\ell \ell $, $\ell h$, $h\ell $, and
$hh$. Although we assume $\epsilon _{g}\tau \ll 1$, it is necessary to
maintain the $\epsilon _{g}\tau $ terms in $\Pi$ as they lead to small
deviations of the eigenvalues of ${\cal V}\Pi $ from unity. Fourier
decomposition in angular momentum space of the disorder line, ${\cal
V}_{\alpha \beta }(\Delta \theta )=\sum_{m}{\cal V}_{\alpha \beta
}^{(m)}e^{im\Delta \theta }$, has the following nonvanishing components:
\begin{eqnarray}
{\cal V}^{(0)} &=&\frac{\left\langle V^{2}\right\rangle }{4}\left(
\begin{array}{cccc}
\;1\; & \;1\; & \;1\; & \;1\; \\
1 & 1 & 1 & 1 \\
1 & 1 & 1 & 1 \\
1 & 1 & 1 & 1
\end{array}
\right) \,, \\
{\cal V}^{(3)} &=&\frac{\left\langle V^{2}\right\rangle }{2}\left(
\begin{array}{cccc}
1 & 0 & 0 & -1 \\
0 & 1 & -1 & 0 \\
0 & -1 & 1 & 0 \\
-1 & 0 & 0 & 1
\end{array}
\right) \,, \\
{\cal V}^{(6)} &=&\frac{\left\langle V^{2}\right\rangle }{4}\left(
\begin{array}{cccc}
1 & -1 & -1 & 1 \\
-1 & 1 & 1 & -1 \\
-1 & 1 & 1 & -1 \\
1 & -1 & -1 & 1
\end{array}
\right) \,.
\end{eqnarray}
We are interested in eigenvalues of $\Pi {\cal V}^{(m)}$ which are
close to unity and their corresponding eigenvectors. Calculating
$C=(1-{\cal V}\Pi )^{-1}{\cal V}$, we find the Cooperon
\begin{eqnarray}  \label{6:cwal}
\lefteqn {C (\Delta \theta) =} \\ \nonumber & & \frac{\left\langle V^2
\right\rangle}{4} \left ( \begin{array}{cccc} 1 & \; 1 \; & \; 1 \; & 1 \\
1 & 1 & 1 & 1 \\ 1 & 1 & 1 & 1 \\ 1 & 1 & 1 & 1 \end{array} \right) \frac
{1}{(Dq^2 - i\omega + \epsilon_{g}^2
\tau / 2) \tau} \\
& & + \, \frac {\left\langle V^2 \right\rangle}{2} \left (
\begin{array}{cccc}
1 & \; 0 \; & \; 0 \; & -1 \\
0 & 0 & 0 & 0 \\
0 & 0 & 0 & 0 \\
-1 & 0 & 0 & 1
\end{array}
\right) \frac {e^{3i \Delta \theta}}{(D q^2 - i \omega) \tau}  \nonumber \\
& & + \, \frac {\left\langle V^2 \right\rangle}{2} \left (
\begin{array}{cccc}
\; 0 \; & 0 & 0 & \; 0 \; \\
0 & 1 & -1 & 0 \\
0 & -1 & 1 & 0 \\
0 & 0 & 0 & 0
\end{array}
\right) \frac {e^{3i \Delta \theta}}{(D q^2 - i \omega + \epsilon_{g}^2
\tau) \tau}
\nonumber \\
& & + \, \frac{\left\langle V^2 \right\rangle}{4} \left (
\begin{array}{cccc}
1 & -1 & -1 & 1 \\
-1 & 1 & 1 & -1 \\
-1 & 1 & 1 & -1 \\
1 & -1 & -1 & 1
\end{array}
\right) \frac {e^{6 i \Delta \theta}}{(Dq^2 - i\omega + \epsilon_{g}^2 \tau
/ 2) \tau} \, .  \nonumber
\end{eqnarray}
Evidently there are two spin-orbit scattering rates, $%
\tau _{{\rm so}}^{-1}=\epsilon _{g}^{2}\tau $ and $\epsilon
_{g}^{2}\tau /2$. When the Cooperon lines are matched in a weak
localization diagram, $\Delta \theta =\pi $. Consequently, a $-1$
term appears in the second and third terms. The $-1$ factor in the
third term is cancelled against another $-1$ factor appearing due
to the contributing matrix elements, $C_{\ell h,h\ell }$ and
$C_{h\ell ,\ell h}$. The first, third and fourth terms hence
generate weak localization while the second term contributes to
weak antilocalization. The emerging picture is identical to the
one obtained for spin-orbit scattering, where spin singlet states
contribute to weak antilocalization while spin triplet states
contribute to weak localization.

The calculation of the quantum interference correction to the
magnetoconductivity follows the standard route (see, e.g.,
Refs.~\onlinecite{HLN80,Averkiev1}), leading to Eq.~(\ref{WALformula}).
This result describes the crossover from weak antilocalization at
$\epsilon_g^2 \tau \tau_\varphi > 1$ to weak localization at
$\epsilon_{g}^2 \tau \tau_\varphi < 1$, as seen in
Fig.~\ref{Fig8(low-mu-MR-for-diff.-P)}.

\subsection{Model and band structure}
\label{6:modelSec}

In this section we discuss the band structure of a two-dimensional hole gas
in a GaAs/AlGaAs heterostructure. This problem was considered theoretically
in refs.~\cite{BroidoSham,Ekenberg}. Here, we present the essential
ingredients, relying on the derivation of ref.~\cite{BroidoSham}. Our main
purpose is to derive the overlap between spinor wavefunctions in the two
bands, which is an important ingredient in the inter-band hole-hole
scattering rate calculation.

The lowest bulk valence bands of GaAs are fourfold degenerate at ${\bf
k}=0$ and split to light and heavy, doubly degenerate bands at finite ${\bf
k}$. When a confining potential in the ${\bf \hat{z}}$ direction is
introduced to create the 2DHG, the degeneracy between the bulk light and
heavy holes is lifted at ${\bf k}=0$ as well. In cases where the confining
potential or the lattice lack inversion symmetry, spin-orbit coupling
further lifts the twofold degeneracy of each band. The spin-orbit coupling
depends on ${\bf k}$ thus at finite ${\bf k}$, the valence band comprises
four bands of different masses and Fermi velocities. Since the splitting
between the bulk light and heavy holes is large, the relevant subbands for
our experiment are the two originating from the bulk heavy band. We denote
these bands as light and heavy but that nomenclature should not be confused
with the light and heavy bulk bands.

In the following calculation we use the spherical approximation \cite{BL73}.
Consider the ${\bf k}\cdot {\bf p}$ Hamiltonian \cite{Lut56} within the
spherical approximation, and in the presence of an asymmetric confining
potential,
\begin{equation}
H=\left( \gamma _{1}+\frac{5\bar{\gamma}}{2}\right) \frac{k^{2}}{2}-\bar{
\gamma}({\bf k}\cdot {\bf J})^{2}+V(z)\,.
\end{equation}
Here, ${\bf J}$ are the angular momentum matrices for $j=3/2$, and $\gamma
_{1}=6.85$ and $\bar{\gamma}=(2\gamma _{2}+3\gamma _{3})/5=2.58$ are the
Luttinger parameters for GaAs in atomic units. The confining potential
$V(z)$ is usually calculated self-consistently within a Hartree
approximation~\cite{BroidoSham,Ekenberg} . Here it is sufficient to assume
that $V(z)$ is the result of such a calculation.

A note regarding our notation: The bulk light and heavy holes are
denoted by capital letters ($L$ and $H$). The light and heavy
bands originating from the bulk heavy band are denote by $\ell $
and $h$. It is the latter two which are relevant for our
discussion.

Within the subspace of light and heavy bulk states, and using a
$J=\frac{3}{2 }$ representation, the two dimensional Hamiltonian takes the
form~\cite{BroidoSham,Ekenberg},
\begin{widetext}
\begin{equation}
H_{\bf k} =  \left(
\begin{array}{cccc}
\displaystyle\epsilon _{{\scriptscriptstyle H}}^{(0)} + \frac{\gamma _{1}
+\bar{ \gamma}}{2}k^{2} & i\sqrt{3}\bar{\gamma}Kk_{-} &
\displaystyle-\frac{\sqrt{3}
}{2}\bar{\gamma}sk_{-}^{2} & 0 \\
-i\sqrt{3}\bar{\gamma}Kk_{+} & \displaystyle\epsilon _{{\scriptscriptstyle
L} }^{(0)}+\frac{\gamma _{1}-\bar{\gamma}}{2}k^{2} & 0 &
\displaystyle-\frac{
\sqrt{3}}{2}\bar{\gamma}sk_{-}^{2} \\
\displaystyle-\frac{\sqrt{3}}{2}\bar{\gamma}sk_{+}^{2} & 0 & \displaystyle
\epsilon _{{\scriptscriptstyle L}}^{(0)}+\frac{\gamma _{1}-\bar{\gamma}}{2}
k^{2} & i\sqrt{3}\bar{\gamma}Kk_{-} \\
0 & \displaystyle-\frac{\sqrt{3}}{2}\bar{\gamma}sk_{+}^{2} &
-i\sqrt{3}\bar{ \gamma}Kk_{+} & \displaystyle\epsilon _{{\scriptscriptstyle
H}}^{(0)}+\frac{ \gamma _{1}+\bar{\gamma}}{2}k^{2}
\end{array}
 \right) .  \nonumber
\end{equation}
\end{widetext}
Here, $k_{\pm } \equiv k_{x}\pm ik_{y}$. The parameters $s \equiv \langle
\psi_\sL^{(0)} | \psi_\sH^{(0)} \rangle$, and $iK \equiv \langle
\psi_\sL^{(0)} | k_{z} | \psi_\sH^{(0)} \rangle $, where $\psi_\sH^{(0)},
\psi_\sL^{(0)}$ are the $ z $-direction states corresponding to the bulk
heavy and light holes, respectively reflect the wave function in the third
dimension. Note $K \neq 0$ for asymmetric confining potential.

To third order in $k$ the Hamiltonian's eigenvalues are
\begin{equation}
\epsilon _{{\bf k}}^{(i)}=\epsilon _{{\scriptscriptstyle H}}^{(0)}+\left(
\frac{\gamma _{1}+\bar{\gamma}}{2}-\frac{3K^{2}\bar{\gamma}^{2}}{\epsilon
_{{ \scriptscriptstyle L}}^{(0)}-\epsilon _{{\scriptscriptstyle H}}^{(0)}}
\right) k^{2}\pm \frac{3Ks\bar{\gamma}^{2}}{\epsilon _{{\scriptscriptstyle
L} }^{(0)}-\epsilon _{{\scriptscriptstyle H}}^{(0)}}k^{3},
\end{equation}
with the upper (lower) sign corresponding to $i=\ell $ ($i=h$).
Within this model the two bands are degenerate in the case of a
symmetric confining potential ($K=0$).

The eigenvectors of $H_{{\bf k}}$ corresponding to the two relevant bands
are $U(\theta )\Xi ^{(\ell )}$ and $U(\theta )\Xi ^{(h)}$, where
\begin{equation}
U(\theta )=\frac{1}{\sqrt{2}}\left(
\begin{array}{cccc}
1 & 0 & 0 & 1 \\
0 & -ie^{-i\theta } & ie^{-i\theta } & 0 \\
0 & 1 & 1 & 0 \\
-ie^{3i\theta } & 0 & 0 & ie^{3i\theta }
\end{array}
\right) \,.
\end{equation}
the angle $\theta $ marks the direction of ${\bf k}$ in the $x$-$y$ plane
(this transformation is a single valued version of the unitary
transformation used in ref.~\onlinecite{BroidoSham}), $\Xi ^{(\ell )}({\bf
k} )=\left( \xi_\sH^{(\ell )} (k), -e^{2 i \theta} \xi_\sL^{(\ell )} (k),
0, 0 \right)$ and $\Xi ^{(h)}({\bf k} ) = \left( 0,0,e^{2i\theta }
\xi_\sL^{(h)} (k), \xi_\sH^{(h)} (k) \right)$. Here $\xi_{\sL, \sH}^{(i)}$
are chosen to be real, and the dependence upon the direction of ${\bf k}$
is given explicitly in the expressions for $ \Xi ^{(i)}({\bf k})$. The $\xi
$'s, depending only on the magnitude of ${\bf k}$, may be explicitly
calculated, but are not needed here. For our calculation we need the
overlap of the spin part of two different eigenstates of the system, as a
function of the two momenta directions. This overlap is given by
\begin{eqnarray} \label{6:overlap}
\lefteqn{M_{ij} (\mathbf{k}, \mathbf{k'}) = \xi^{\dagger (i)} (\mathbf{k})
U^\dagger (\theta) U (\theta') \xi^{(j)} (\mathbf{k'})} \\
&&  \nonumber \\
&& \!\!\!\!\!\!\! =\left\{
\begin{array}{lcl}
\displaystyle \frac{1}{2} \left( 1+e^{3i\Delta \theta } \right)
\xi_\sH^{(i)}(k) \xi_\sH^{(i)}(k^{\prime }) & & \\
\displaystyle \;\;\;\;\;\;\;\;\;\;\;\;\;\; + \frac{1}{2}\left( e^{i\Delta
\theta }+e^{2i\Delta \theta }\right) \xi_\sL^{(i)}(k) \xi_\sL^{(i)} (k')
& \; & i=j, \\ &  &  \\
\displaystyle \frac{1}{2} \left( 1-e^{3i\Delta \theta }\right) \xi_\sH^{(i)}
(k) \xi_\sH^{(j)}(k') & & \\
\displaystyle \;\;\;\;\;\;\;\;\;\;\;\;\;\; + \frac{1}{2} \left( e^{i\Delta
\theta }-e^{2i\Delta \theta }\right) \xi_\sL^{(i)} (k) \xi_\sL^{(j)} (k') &
& i\neq j,
\end{array}
\right.  \nonumber
\end{eqnarray}
where $\Delta \theta = \theta' - \theta $, and the indices $i$ and $j$ take
the value $\ell $ or $h$. This result is used in the calculation of the
transport properties in Section (\ref{6:calcSec}).

\begin{acknowledgments}
We thank I.~L. Aleiner, B.~L. Altshuler, A.~M. Finkel'stein, B.~I.
Halperin, A. Kamenev, D.~L. Maslov, B.~N. Narozhny, A. Punnoose, and M.
Reznikov for stimulating discussions. This work was supported by the
Israeli science foundation and by German-Israeli DIP grant.
\end{acknowledgments}

\appendix

\section{}

\label{appendix}

\begin{figure}
\begin{center}
\setlength{\unitlength}{3.4in}
\begin{picture}(1,0.69)
\put(0.05,0){\includegraphics[width=0.95\unitlength]{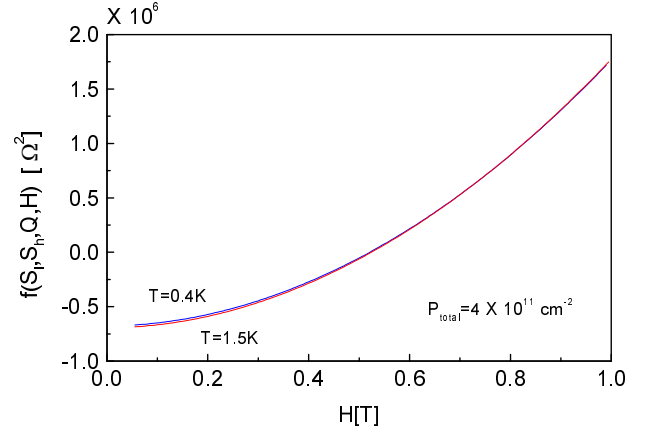}}
\end{picture}
\end{center}
\caption{$f(S_{l},S_{h},Q,H)$ vs.~magnetic field at two different
temperatures for $p_{total}=4\times 10^{11}cm^{-2}$.}
\label{fig20(f(S_l,S_h,Q,H) vs. H)}
\end{figure}

The function $f(S_{l},S_{h},Q,H)$ in Eq.~(\ref{MR(eq.)}) is given by
\begin{eqnarray*}
f(S_{l},S_{h},Q,H) &=&\frac{1}{[1+(H/W)^{2}]^{2}}\cdot \frac{1}{
(S_{l}+S_{h}+2Q)^{4}}\cdot \\
&&\left[ C_{0}+C_{1}\cdot H^{2}+C_{2}\cdot H^{4}+C_{3}\cdot
H^{6}\right] ,
\end{eqnarray*}
where
\begin{eqnarray*}
C_{0} &=&-(S_{l}+S_{h}+2Q)^{2}\cdot (S_{l}S_{h}-Q^{2})^{2}, \\
C_{1} &=&Q_{0}+Q\cdot Q_{1}+Q^{2}\cdot Q_{2}+Q^{3}\cdot
Q_{3}+Q^{4}\cdot Q_{4},
\\
C_{2} &=&P_{0}+Q\cdot P_{1}+Q^{2}\cdot P_{2}, \\
C_{3} &=&R_{l}^{2}R_{h}^{2}(R_{l}+R_{h})^{2},
\end{eqnarray*}
and
\begin{eqnarray*}
Q_{0} & \! = \! & (R_{h}S_{l}^{2}+R_{l}S_{h}^{2})^{2}-2(S_{l} \! + \!
S_{h})(R_{h}^{2}S_{l}+R_{l}^{2}S_{h})S_{l}S_{h},
\\
Q_{1} & \! = \! & 4 \left[ (S_{l} \! - \! S_{h}) (R_{h}^{2}S_{l}^{2} \! -
\! R_{l}^{2}S_{h}^{2}) \!+ \! 2R_{l}R_{h}S_{l}S_{h} (S_{l} \! + \! S_{h})
\right] \! ,
\\
Q_{2} & \! = \! & 2 \left[ S_{l}S_{h}(R_{l} + R_{h})^{2} +
R_{l}R_{h}(S_{l}-S_{h})^{2} \right. \\ & & \left. + \,
4(R_{h}S_{l}+R_{l}S_{h})^{2} \right] \! , \\
Q_{3} & \! = \! & 8(R_{h}^{2}S_{l}+R_{l}^{2}S_{h}), \\
Q_{4} & \! = \! & R_{l}^{2}-6R_{l}R_{h}+R_{h}^{2}, \\
P_{0} & \! = \! & 2R_{l}^{2}R_{h}^{2}(S_{l}-S_{h})^{2} -
(R_{h}^{2}S_{l}-R_{l}^{2}S_{h})^{2} \\ & & + \,
2R_{l}R_{h}(R_{h}^{2}S_{l}^{2}+R_{l}^{2}S_{h}^{2}),
\\
P_{1} & \! = \! & 4R_{l}R_{h}[R_{h}S_{l}(R_{l}+2R_{h}) + R_{l}S_{h}(R_{h}+2R_{l})], \\
P_{2} & \! = \! & 2R_{l}R_{h}(R_{h}^{2}+R_{l}^{2}).
\end{eqnarray*}
Fig.~\ref{fig20(f(S_l,S_h,Q,H) vs. H)} depicts $f(S_{l},S_{h},Q,H)$ vs.
magnetic field at $T=0.4$ and $1.5K$ for $p_{total}=4\times
10^{11}cm^{-2}$.


\begin{thebibliography}{99}
\bibitem{Kravchenko94}  S. V. Kravchenko, G. V. Kravchenko, J. E. Furneaux,
V. M. Pudalov, and M. D'Iorio, Phys. Rev. B {\bf 50}, 8039 (1994).

\bibitem{gang4}  E. Abrahams, P. W. Anderson, D. C. Licciardello, and T. V.
Ramakrishnan, Phys. Rev. Lett. {\bf 42}, 673 (1979).

\bibitem{metal}  A. M. Finkel'stein, Zh. Eksp. Teor. Fiz. {\bf 84}, 168
(1983) [Sov. Phys.-JETP {\bf 57}, 97 (1983)]; A. M. Finkel'stein, Z. Phys.
A {\bf 56}, 189 (1984); C. Castellani {\em et al.}, Phys. Rev. B {\bf 30},
1596 (1984); C. Castellani, C. DiCastro, P. A. Lee, and M. Ma, Phys. Rev. B
{\bf 30}, 527 (1984); C. Castellani, C. Di Castro, and P. A. Lee, Phys.
Rev. B {\bf 57}, R9381 (1998).

\bibitem{silicon}  S. V. Kravchenko, W. E. Mason, G. E. Bowker, J. E.
Furneaux, V. M. Pudalov, and M. D'Iorio, Phys. Rev. B {\bf 51}, 7038
(1995); S. V. Kravchenko, D. Simonian, M.P. Sarachik, W. Mason, and J. E.
Furneaux, Phys. Rev. Lett. {\bf 77}, 4938 (1996); D. Popovi\'{c}, A. B.
Fowler, and S. Washburn, Phys. Rev. Lett. {\bf 79}, 1543 (1997); R.
Heemskerk and T. M. Klapwijk, Phys. Rev. B {\bf 58}, R7154 (1998).

\bibitem{SiGe}  P. T. Coleridge, R. L. Williams, Y. Feng, and P. Zawadzki,
Phys. Rev. B {\bf 56}, R12764 (1997); J. Lam, M. D'Iorio, D. Brown, and H.
Lafontaine, Phys. Rev. B {\bf 56}, R12741 (1997).

\bibitem{AlAs}  S. J. Papadakis and M. Shayegan, Phys. Rev. B {\bf 57}, R15068
(1998).

\bibitem{2DEGGaAs}  Y. Hanein, D. Shahar, J. Yoon, C. C. Li, D. C. Tsui,
and H. Shtrikman, Phys. Rev. B {\bf 58}, R13338 (1998).

\bibitem{InAs}  E. Ribeiro, R. D. J\"{a}ggi, T. Heinzel, K. Ensslin, G.
Medeiros-Ribeiro, and P. M. Petroff, Phys. Rev. Lett. {\bf 82}, 996 (1999).

\bibitem{Hanein1}  Y. Hanein, U. Meirav, D. Shahar, C. C. Li, D. C. Tsui and
H. Shtrikman, Phys. Rev. Lett. {\bf 80}, 1288 (1998).

\bibitem{Pepper1}  M. Y. Simmons, A. R. Hamilton, M. Pepper, E. H. Linfield,
P. D. Rose and D. A. Ritchie, Phys. Rev. Lett. {\bf 80} 1292 (1998).

\bibitem{Papadakis1}  S. J. Papadakis, E. P. De Poortere, H. C. Manoharan,
M. Shayegan, and R. Winkler, Science {\bf 283}, 2056 (1999).

\bibitem{Sivan}  Y. Yaish, O. Prus, E. Buchstab, S. Shapira, G. Ben Yosef,
U. Sivan, and A. Stern, Phys. Rev. Lett. {\bf 84}, 4954 (2000).

\bibitem{superconductivity}  P. Phillips, Y. Wan, I. Martin, S. Kngsh, and
D. Dalidovich, Nature {\bf 395}, 253 (1998); D. Belitz and T. R.
Kirkpatrick, Phys. Rev. B {\bf 58}, 8214 (1998).

\bibitem{Varma}  Q. Si and C. M. Varma, Phys. Rev. Lett. {\bf 81}, 4951
(1998).

\bibitem{Wignerglass}  S. Chakravarty, S. Kivelson, C. Nayak, and K.
V\"{o}lker, Phil. Mag. B {\bf 79}, 859 (1999).

\bibitem{scaling}  V. Dobrosavljevi\'{c}, E. Abrahams, E. Miranda, and S.
Chakravarty, Phys. Rev. Lett. {\bf 79}, 455 (1997); D. Simonian, S. V.
Kravchenko, M. P. Sarachik,  and V. M. Pudalov, Phys. Rev. B {\bf 57},
R9420 (1998).

\bibitem{Hamilton}  A. R. Hamilton, M. Y. Simmons, M. Pepper, E. H.
Linfield, P. D. Rose and D. A. Ritchie, Phys. Rev. Lett. {\bf 82}, 1542
(1999).

\bibitem{Savchenko}  S. S. Safonov, S. H. Roshko, A. K. Savchenko, A. G.
Pogosov, and Z. D. Kvon, Phys. Rev. Lett. {\bf 86}, 272 (2001).

\bibitem{Papadakis2}  S. J. Papadakis, E. P. De Poortere, M. Shayegan, and
R. Winkler, Phys. Rev. Lett. {\bf 84}, 5592 (2000); S. J. Papadakis, E. P.
De Poortere, and M. Shayegan, Phys. Rev. B {\bf 62}, 15373 (2000).

\bibitem{Prus}  O. Prus {\em et al.,} (2000) unpublished.

\bibitem{Popovich}  X. G. Feng, D. Popovi\'{c}, and S. Washburn, Phys. Rev.
Lett. {\bf 83}, 368 (1999)

\bibitem{Newliquidphase}  S. He and X. C. Xie, Phys. Rev. Lett. {\bf 80},
3324 (1998).

\bibitem{maslovaltshuler}  B. L. Altshuler and D. Maslov, Phys. Rev. Lett.
{\bf 82}, 145 (1999).

\bibitem{Das Sarma}  A. Gold and V. T. Dolgopolov, Phys. Rev. B {\bf 33},
1076 (1986); S. Das Sarma and E. H. Hwang, Phys. Rev. Lett. {\bf 83}, 164
(1999).

\bibitem{Hamilton2}  A. R. Hamilton, M. Y. Simmons, M. Pepper and D. A.
Ritchie, Aust. J. Phys. {\bf 53}, 523 (2000).

\bibitem{Pudalov}  V. M. Pudalov, Pis'ma Zh. Eksp. Teor. Fiz. {\bf 66}, 168
(1997).

\bibitem{Meir}  Y. Meir, Phys. Rev. Lett. {\bf 83}, 3506 (1999); Y. Meir,
Phys. Rev. B {\bf 61, }16470 (2000).

\bibitem{AshcroftMermin}  N. W. Ashcroft and D. D. Mermin, {\em Solid State
Physics}, Harcourt Brace College Publishers (1976), p. 240.

\bibitem{Gantmakher}  V. F. Gantmakher and Y.B. Levinson, Sov. Phys. JETP,
{\bf 47}, 133 (1978).

\bibitem{Zaremba}  E. Zaremba, Phys. Rev. B {\bf 45}, 14143 (1992).

\bibitem{Stormer}  H. L. Stormer, Z. Schlesinger, A. Chang, D. C. Tsui, A.
C. Gossard, and W. Wiegmann, Phys. Rev. Lett. {\bf 51}, 126 (1983).

\bibitem{Eisenstein}  J. P. Eisenstein, H. L. Stormer, V. Narayanamurti, A.
C. Gossard, and W. Wiegmann, Phys. Rev. Lett. {\bf 53}, 2579 (1984).

\bibitem{Shayegan}  J. P. Lu, J. B. Yau, S. P. Shukla, M. Shayegan, L.
Wissinger, U. R\"{o}ssler, and R. Winkler, Phys. Rev. Lett. {\bf 81}, 1282
(1998).

\bibitem{BroidoSham}  D. A. Broido and L. J. Sham, Phys. Rev. B {\bf 31},
888 (1985).

\bibitem{Ekenberg}  U. Ekenberg and M. Altarelli, Phys. Rev. B {\bf 32}, 3721
(1985).

\bibitem{Goldoni}  G. Goldoni and F. M. Peeters, Phys. Rev. B {\bf 51}, 17806
(1995).

\bibitem{Winkler}  R. Winkler, S. J. Papadakis, E. P. De Poortere, and M.
Shayegan, Phys. Rev. Lett. {\bf 84}, 713 (2000); R.Winkler, cond-mat/0002003.

\bibitem{Murzin}  S. S. Murzin, S. I. Dorozhkin, G. Landwehr, and A. C.
Gossard, JETP Lett. {\bf 67}, 113 (1998).

\bibitem{EEIreview}  For a review, see B. L. Altshuler and A. G. Aronov,
\textit{Electron Electron Interactions in Disordered Systems},
edited by M. Pollak and A. L. Efros (North Holland, Amsterdam,
1984).

\bibitem{Paalanen}  M. A. Paalanen, D. C. Tsui, and J. C. M. Hwang, Phys.
Rev. Lett. {\bf 51}, 2226 (1983).

\bibitem{Choi}  K. K. Choi, D. C. Tsui, and S. C. Palmateer, Phys. Rev. B
{\bf 33}, 8216 (1986).

\bibitem{Hanein2}  Y. Hanein, D. Shahar, J. Yoon, C. C. Li, D. C. Tusi and
H. Shtrikman, Phys. Rev. B {\bf 58}, R7520 (1998).

\bibitem{Pepper2}  M. Y. Simmons, A. R. Hamilton, M. Pepper, E. H. Linfield,
P. D. Rose and D. A. Ritchie, Phys. Rev. Lett. {\bf 84}, 2489 (2000).

\bibitem{Dresselhaus}  P. D. Dresselhaus, C. M. A. Papavassiliou, R. G.
Wheeler and R. N. Sacks, Phys. Rev. Lett. {\bf 68}, 106 (1992).

\bibitem{Knap}  W. Knap, C. Skierbiszewski, A. Zduniak, E.
Litwin-Staszewska, D. Bertho, F. Kobbi, J. L. Robert, G. E., Pikus, F. G.
Pikus, S. V. Iordanskii, V. Mosser, K. Zekentes, and Yu. B. Lyanda-Geller,
Phys. Rev. B {\bf 53}, 3912 (1996).

\bibitem{Lyanda-Geller}  Y. Lyanda-Geller, Phys. Rev. Lett. {\bf 80}, 4273
(1998).

\bibitem{Averkiev1}  N. S. Averkiev, L. E. Golub and G. E. Pikus, JETP. {\bf
86}, 780 (1998).

\bibitem{Dyakonov}  M. I. D'yakonov and V. I. Perel', Zh. Eksp. Teor. Fiz.
{\bf 60}, 1954 (1971) [Sov. Phys. JETP {\bf 33}, 1053 (1971)].

\bibitem{Pedersen1}  S. Pedersen, C. B. Sorensen, A. Kristesen, P. E.
Lindelof, L. E. Golub and N. S. Averkiev, Phys. Rev. {\bf 60}, 4880 (1999).

\bibitem{Fukuyama2}  H. Fukuyama, J. Phys. Soc. Jpn. {\bf 50}, 3407 (1981).

\bibitem{Finkel'stein}  A. M. Finkel'stein, Zh. Eksp. Teor. Fiz. {\bf 84},
168 (1983) [Sov. Phys.-JETP {\bf 57}, 97 (1983)].

\bibitem{Fukuyama1}  H. Fukuyama, Y. Isawa, and H. Yasuhara, J. Phys. Soc.
Jpn. {\bf 52}, 16 (1983).

\bibitem{Lee}  P. A. Lee and T. V. Ramakrishnan, Phys. Rev. B{\bf 26}, 4009
(1982).

\bibitem{Altshuler}  B. L. Altshuler, A. G. Aronov, A. I. Larkin, and D. E.
Khmelnitskii, Zh. Eksp. Teor. Fiz. {\bf 81}, 768 (1981) [Sov. Phys.-JETP
{\bf 54}, 411 (1981)].

\bibitem{Bergman}  G. Bergmann, Phys. Rev. B {\bf 25}, 2937 (1982);
Phys. Rev. Lett. {\bf 48}, 1046 (1982); Z. Phys. B {\bf 48}, 5 (1982).

\bibitem{Davies}  R. A. Davies and M. Pepper, J. Phys. C {\bf 16}, L353
(1983).

\bibitem{Brunthaler}  G. Brunthaler, A. Prinz, G. Bauer, V. M. Pudalov, E.
M. Dizhur, J. Jaroszynski, P. Glod and T. Dietl, Ann. Phys.-Berlin {\bf 8},
579 (1999).

\bibitem{Coleridge2}  P. T. Coleridge, A. S. Sachrajda and P. Zawadzki,
cond-mat/9912041.

\bibitem{Meirav}  U. Meirav, M. Heiblum and F. Stern, Appl. Phys. Lett. B
{\bf 52}, 1268 (1988).

\bibitem{Dultz}  S. C. Dultz and H. W. Jiang, Phys. Rev. Lett. {\bf 84},
4689 (2000).

\bibitem{paper3}  Y. Yaish, O. Prus, E. Buchstab, G. Ben Yosef, and U.
Sivan, unpublished (2001); Y. Yaish, Ph.D. Thesis, Technion-IIT
(2001) pp. 88-95.

\bibitem{Yacoby}  S. Ilani, A. Yacoby, D. Mahalu and H. Shtrikman, Phys.
Rev. Lett. {\bf 84}, 3133 (2000).

\bibitem{Narozhny}  G. Zala, B.~N. Narozhny, and I.~L. Aleiner,
cond-mat/0105406. See also Y.~Y. Proskuryakov, A.~K. Savchenko, S.~S.
Safonov, M. Pepper, M.~Y. Simmons, and D.~A. Ritchie, cond-mat/0109261.

\bibitem{FH}  K. Flensberg and B.~Y.-K. Hu, Phys. Rev. Lett. {\bf 73}, 3572
(1994); Phys. Rev. B {\bf 52}, 14796 (1995).

\bibitem{Yaib}  Y. Yaish, unpublished (2001).

\bibitem{HLN80}  S. Hikami, A.~I. Larkin, and Y. Nagaoka, Prog. Theor. Phys.
{\bf 63}, 707 (1980).

\bibitem{RB85}  D. Rainer and G. Bergmann, Phys. Rev. B {\bf 32}, 3522
(1985).

\bibitem{GDK98b}  I.~V. Gornyi, A.~P. Dmitriev, and {V.~Yu.\ Kachorovskii},
Pis'ma Zh. \'Eksp. Teor. Fiz. {\bf 68}, 314 (1998) [JETP Lett. {\bf 68}, 338
(1998)].

\bibitem{BL73}  A. Baldereschi and N.~O. Lipari, Phys. Rev. B {\bf 8}, 2697
(1973).

\bibitem{Lut56}  J.~M. Luttinger, Phys. Rev. {\bf 102}, 1030 (1956).
\end{thebibliography}
\end{document}